\documentclass[aps,prb,amsmath,superscriptaddress,amssymb,longbibliography,twocolumn]{revtex4-2}

\usepackage{amsmath}
\usepackage{graphicx}% Include figure files
\graphicspath{ {images/}}
\usepackage{dcolumn}% Align table columns on decimal point
\usepackage{bm}% bold math
\usepackage[	
citecolor=magenta,
colorlinks,
linkcolor=magenta,
urlcolor=magenta,
]{hyperref}
\usepackage{braket}
\usepackage{appendix}
\usepackage{wasysym}
\usepackage{slashed}
\usepackage[table, svgnames, dvipsnames]{xcolor}
\usepackage{soul}
\usepackage{graphicx}
\usepackage{subfigure}

\usepackage{caption}
\usepackage{ragged2e}
\DeclareCaptionJustification{justified}{\justifying}
\newcommand{\mcl}[1]{\mathcal{#1}}
\newcommand{\mbf}[1]{\mathbf{#1}}

\DeclareUnicodeCharacter{2212}{-}

\begin{document}

\title{Ab-initio Insights on the Fermiology of \texorpdfstring{$d^1$}{} Transition metals in Honeycomb lattice : Hierarchy of hopping pathways and spin-orbit coupling}
\author{Manoj Gupta$^{a}$}
\thanks{$a$ The authors contributed equally to this work.}
 \email{gpta.mnj@gmail.com}
 \affiliation{Department of Condensed Matter Physics and Materials Science,
S. N. Bose National Centre for Basic Sciences, Kolkata 700098, India}
 \author{Basudeb Mondal$^{a}$}
 \email{basudeb.mondal@icts.res.in}
 \affiliation{International Centre for Theoretical Sciences, Tata Institute of Fundamental Research, Bengaluru 560 089, India}
  \author{Subhro Bhattacharjee}
 \email{subhro@icts.res.in}
 \affiliation{International Centre for Theoretical Sciences, Tata Institute of Fundamental Research, Bengaluru 560 089, India}
  \author{Tanusri Saha Dasgupta}
 \email{t.sahadasgupta@gmail.com}
 \affiliation{Department of Condensed Matter Physics and Materials Science,
S. N. Bose National Centre for Basic Sciences, Kolkata 700098, India}
\date{\today}

\begin{abstract}
Motivated by the intriguing suggestion of realizing SU(8) Dirac semi-metal with $J=3/2$ electrons on a honeycomb lattice, we provide a systematic study of the interplay of various hopping pathways and atomic spin-orbit coupling for the low energy electrons in candidate d$^1$ transition metal halides MX$_3$ (M=Ti, Zr, Hf; X=F, Cl, Br). By combining first principle calculations and minimal hopping Hamiltonian, we uncover the role of dominant direct metal-metal hopping on top of indirect  metal-halide-metal hopping. This sets up a hierarchy of hopping pathways that centrally modify the SU(8) picture for the above materials. These hopping interactions, along with the spin-orbit coupling, lead to a plethora of exactly compensated metals instead of the SU(8) Dirac semi-metal.  Remarkably the same can be understood as descendants of a topological insulator obtained by gapping out the SU(8) Dirac semi-metallic phase. The resultant compensated metals have varied Fermi surface topology and are separated by Lifshitz phase transitions. We discuss the implications of the proximate Lifshitz transition, which may be accessed via strain, in the context of the relevant materials.
\end{abstract}

%%%%%%%%%%%%%%%%%%%%%%%%%%
\maketitle
%%%%%%%%%%%%%%%%%%%%%%%%%%

%\tableofcontents
%%%%%%%%%%%%%%%%%%%%%%%%%%
\section{Introduction}

Electronic phases of transition metal compounds with active $d$-electrons pose some of the outstanding problems in condensed matter physics. In addition to the well-known effect of electron-electron interactions, in recent times it has been realized that the atomic spin-orbit coupling (SOC) plays an important role in shaping up the structure of the low energy theory of the 4d and 5d transition metal compounds~\cite{cao2013frontiers,witczak2013correlated}. This provides the scope to study, design and engineer newer platforms of quantum materials supporting novel electronic phases resulting from the interplay of quantum entanglement and symmetries as evidenced in spin–orbit assisted Mott insulators, quantum spin liquids, excitonic magnetism, multipolar orderings, and correlated topological semimetals~\cite{takayama2021spin,wan2011topological,RevModPhys.82.3045,RevModPhys.83.1057,chen2009experimental,PhysRevLett.118.217202,PhysRevLett.102.017205,PhysRevB.82.174440,broholm2020quantum,knolle2019field,takagi2019concept}.

In particular, for 4d and 5d transition metals in octahedral crystal field and with active $t_{2g}$ orbitals, the strong SOC can split the atomic $t_{2g}$ orbitals into effective higher energy $J=1/2$ doublet and lower energy $J=3/2$ quadruplet~\cite{takayama2021spin}. In this regard, the physics of the $J=1/2$ orbitals has been studied extensively~\cite{kim2008novel,kim2009phase,kim2012magnetic}. In comparison, the compounds with active $J=3/2$ manifold have received somewhat less attention in spite of the potential to harbor equally or even richer low energy physics. 

A potential candidate of expected $J=3/2$ physics are the transition metal compounds in $d^{1}$ electronic state where the $t_{2g}$ orbitals are at 1/6th filling. At the strong SOC limit, this leads to an empty $J=1/2$ manifold, while the $J=3/2$ orbitals are quarter filled. Among these compounds, of particular interest to us are layered transition metal (M) halides (X), of general formula MX$_3$~\cite{cryst7050121}. The $3^{+}$ transition metal (M) cations in these compounds, in the octahedral setting of $1^{-1}$ halide anions form honeycomb lattice by edge sharing of MX$_6$ octahedra. Yamada {\it et. al.}~\cite{PhysRevLett.121.097201} proposed ZrCl$_3$, a member of these the honeycomb $d^1$ family, as candidate material exhibiting SU(4) symmetric spin Hamiltonian in the strong coupling limit. Mondal {\it et. al.}~\cite{mondal2023emergent} showed that the same can lead to SU(8) Dirac semi-metal (DSM) in the non-interacting limit. 

It is to be noted though, the above Hamiltonian(s) were derived under the assumption of infinite SOC limit by projecting to the $J=3/2$ orbitals, and by considering only indirect hopping, {\it i.e.} hopping between two M sites via the X. As expected, both of these assumptions are idealized limits vis-a-vis the candidate materials. There are other microscopic energy-scales even at the single electron level -- the various hopping amplitudes and non-cubic crystal field splitting -- that can compete and change the low energy physics, and hence needs to be understood. While the SU(4) or SU(8) may provide an interesting starting point to capture the intermediate coupling physics in real materials, generically one expects that these other energy scales would reduce the fine tuned SU(4) or SU(8) symmetries for the spin model or the free Dirac theory, respectively. What then is the resultant structure of the low energy theory and the possible electronic phases in this family of materials ? More interestingly, the details of these competing microscopic interactions can change across the periodic table even within the same class of  MX$_3$ materials, {\it e.g.}, in moving from 3d to 4d to 5d transition metal, (M=Ti $\rightarrow$ Zr $\rightarrow$ Hf) and in moving from 2p to 3p to 4p halogen (X=F $\rightarrow$ Cl $\rightarrow$ Br). 

In this paper, we present {\it ab-initio} Density functional theory (DFT) insights into the estimates of the different hopping pathways and their effect on the 
low energy minimal  hopping models to reveal a rich structure of the possible phases relevant to MX$_3$. Guided by our DFT calculations, we conclude that the low energy fermiology of the electrons requires SOC and a minimal set of four hopping pathways that include two direct $dd$ hopping, $t_{dd\sigma}$ and $t_{dd\pi}$, in addition to two well known~\cite{kanamori1959superexchange} indirect ones via the intermediate halide ion, $t_{ddm}$, $t_{ddm^{'}}$. Across the transition metal series, the hierarchy of direct and indirect hopping shows an interesting evolution with the fluorides being notably different from the chlorides and bromides. This results in a marked deviation from the indirect hopping only model at large SOC.

Starting with the SU(8) DSM, we provide a controlled understanding of the effect of the other hopping pathways as well as finite SOC. Our DFT results suggest that the largest of the microscopic energy scales is the direct overlap between the transition metals ions, $t_{dd\sigma}$. This, along with finite SOC ($\lambda$), allowing for the mixing amongst the different $t_{2g}$ orbitals provide a natural setting to explore the phase diagram in the $t_{dd\sigma}/t_{ddm}-\lambda$ plane around the SU(8) semi-metal. For chlorides and bromides, the effect of both these perturbations can be understood in terms of gapping out of the SU(8) Dirac fermions. However, while finite $t_{dd\sigma}$ gives a $Z_2$ free fermion symmetry protected topological insulator (SPT)~\cite{PhysRevLett.95.226801,Senthil_SPT}, finite $\lambda$ leads to a trivial one with an intervening SU(2) DSM line separating the two. The primary effect of sub-leading direct and indirect hopping, $t_{dd\pi}$ and $t_{ddm'}$ is to change the details of the band structure around the Fermi level to give rise to particle-hole pockets resulting in a plethora of exactly compensated metals with varied Fermi surface topology. Two characteristic features of the Fermi surfaces relevant for the materials, we note, are -- (a) near nesting of different sections of the Fermi pockets, and, (b) proximity to Lifshitz transition which leads to the change in the Fermi surface topology. Remarkably, all these compensated metals are direct descendants of  the topological insulator obtained in the SU(8) limit and carry a non-trivial $Z_2$ index. Using our ab-initio estimates of the band parameters, we place various materials (chlorides and bromides) on the relevant part of the $\lambda$-hopping phase diagram. Interestingly, we find that the chlorides lie close to the phase boundary between the different compensated metals separated by a Lifshitz transition. This opens up the possibility of strain induced Lifshitz transition in monolayer MCl$_3$.   

The rest of the paper is organized as follows. In Sec. \ref{method} we introduce the DFT computation details that we use to obtain the electronic band structure, as well as the construction of the low energy Hamiltonian of the compounds. In Sec. \ref{sec_structure}, the crystal structure of the nine compounds, MX$_3$ where M=Ti, Zr, Hf and X=F, Cl, Br are discussed. This is followed by the discussion of DFT band structure in Sec. \ref{band-structure} and DFT-derived low energy tight binding model in Sec.\ref{TB}. The phase diagrams of the low-energy hopping model in SOC strength and hopping space upon systematic introduction of hopping have been discussed in subsequent section and sub-sections (cf Sec.\ref{phase-diagram}). This also includes the phase diagram relevant to the discussed chloride and bromide compounds and discussion on the possible  Lifshitz transition. We close the result section with a brief overview on the consequences of parameters relevant for fluorides in the  $t_{dd\sigma}/t_{ddm}-\lambda$ phase diagram in Sec.\ref{sec_flourides}. In Sec.\ref{sec_summary} summary and outlook are presented. Supporting technical details are given in SM~\cite{SM}.

%%%%%%%%%%%%%%%%%%%%%

%%%%%%%%%%%%%%%%%%%%%%%
\section{Methods and computational Details}
\label{method}
The first-principles DFT calculations were carried out using plane-wave basis and projector augmented-wave potential~\cite{PhysRevB.50.17953,TACKETT2001348,10.1063/1.1926272}, as implemented in Vienna Abinitio Simulation package~\cite{kresse1993ab,eliav1994open,kresse1999ultrasoft}. The Perdew-Burke-Ernzerhof generalized gradient approximation (GGA)~\cite{perdew1996generalized} was used to approximate the exchange-correlation functional. To check the influence of correlation effect
at transition metal site, beyond GGA, GGA+U with supplemented Hubbard U correction was carried out~\cite{dudarev1998electron}. Further, to handle the van Waals interaction between the layers, dispersion corrected GGA+U+D2 functional was used~\cite{grimme2006semiempirical}. The convergence of energies and forces, was ensured by using a plane-wave energy cutoff of 600 eV and BZ sampling with 6 $\times$ 6 $\times$ 6 Monkhorst-Pack grids. During the structural relaxation, the ions were allowed to move until the atomic forces became lower than 0.0001 eV/Å.

The construction of DFT-derived low energy, few band Hamiltonian in the effective $t_{2g}$ Wannier basis of the transition metal ions was achieved through downfolding technique of integrating out degrees of freedom that are 
not of interest, starting from the all-orbital DFT bandstructure, calculated in muffin-tin orbital (MTO) basis. The self-consistent potentials required for
these calculations, were generated through Stuttgart implementation of Linear-MTO (LMTO) package~\cite{andersen1984explicit}, while the downfolding calculations were performed in Nth-order muffin-tin orbital (NMTO) basis~\cite{andersen2000muffin}. For muffin-tin orbital calculations, the metal atom-centred MT radii were chosen to be in the range of 1.30-1.53 {\AA}, 1.47-1.66 {\AA}, and 1.44-1.70 {\AA} for Ti, Zr, and Hf, respectively. MT radii of 
0.94-0.97 {\AA}, 1.33-1.34 {\AA}, and 1.46-1.48 {\AA} were chosen 
for halogen atoms, F, Cl, and Br, respectively.
The consistency of the results
between plane-wave and muffin-tin orbital basis was checked in terms of bandstructure and density of states.

%%%%%%%%%%%%%%%%%%%%%%%%%%%%%%

\section{Crystal Structure and $d$ Level Splitting of the Studied Compounds}
\label{sec_structure}

\begin{figure*}
\includegraphics[scale=0.6]{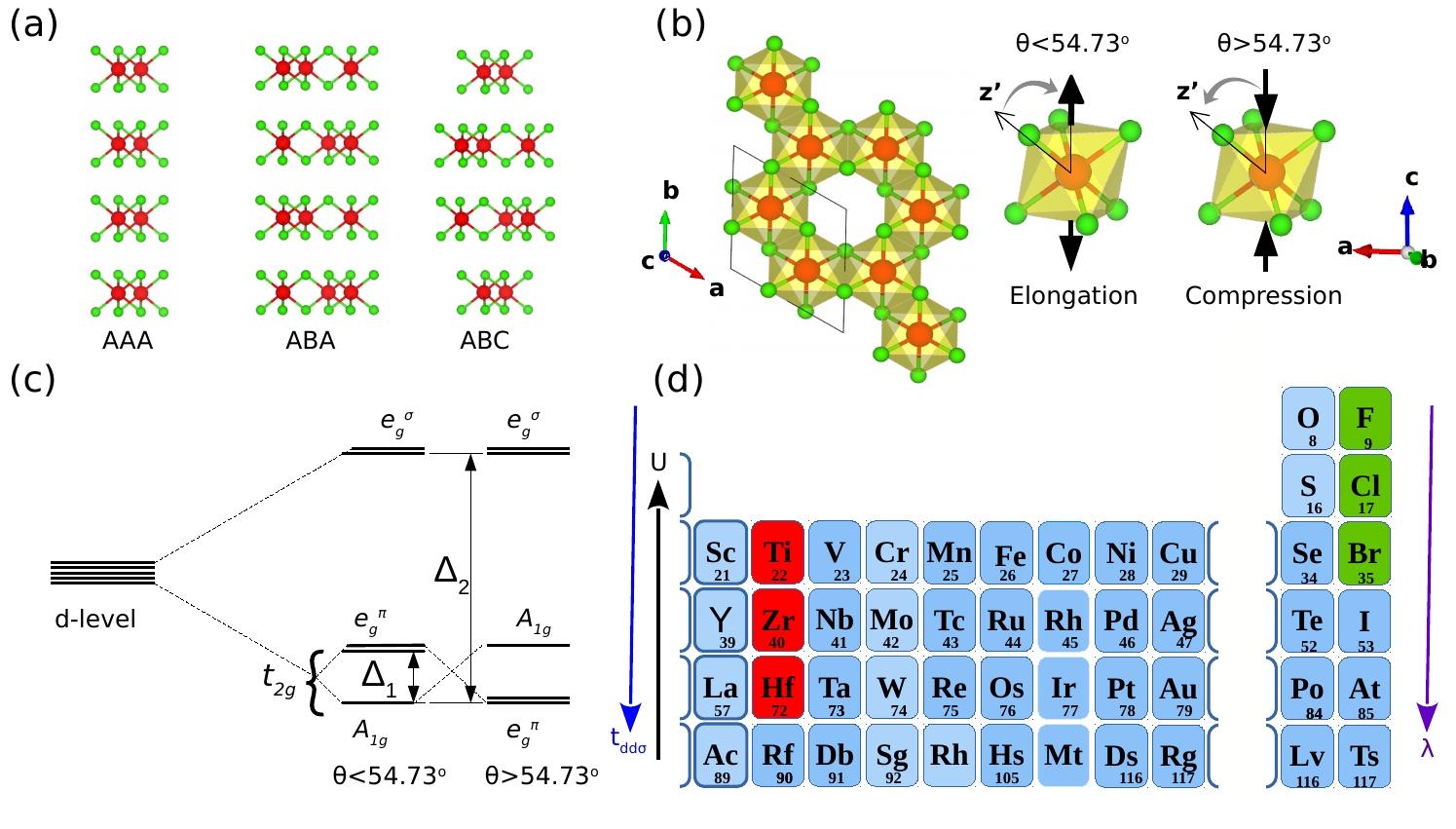}
\caption{(a) Different possible stacking of the honeycomb layers in MX$_3$ family. The M and X atoms
are shown as red and green colored balls respectively. (b) Left panel shows the honeycomb layer of M atoms formed by the edge-sharing of MX$_6$ octahedra. Right panel shows the trigonal distortion of the MX$_6$ octahedra. (c) The octahedral ($\Delta_2$) and trigonal ($\Delta_1$) crystal field splitting of the $d$ levels of M. (d) The choice of M (marked in red) and X (marked in green) elements of the compounds under study. The variation of the three primary energy scales of the problem -- SOC ($\lambda$), orbital size dictating the direct metal-metal hopping (t$_{dd\sigma}$) and Coulomb interaction (U) is shown.}
\label{fig:stacking}
\end{figure*}

The layered structure of MX$_3$ compounds comprises of two-dimensional hexagonal nets of MX$_3$, van der Waals stacked on each other. Clearly, there can be many variants on the stacking sequence in these layered materials. Due to the weak van der Waals interactions between layers, different stacking sequences result only in small energy differences. For example, TiCl$_3$ is reported to adopt ABA, AAA, and ABC stacking sequences, as shown in Figure \ref{fig:stacking}(a), with ABA stacking resulting in 
P$\bar{3}$1c~\cite{troyanov1991x}, AAA stacking in  P$\bar{3}$1m ~\cite{allegra1962calcolo}, and ABC stacking in R$\bar{3}$~\cite{troyanov1991x} space groups. Though these polymorphs have different crystal space groups, their in-plane geometry as well as inter-layer distance show little variation, less than 0.2 $\%$ for the specific case of TiCl$_3$~\footnote{The in-plane lattice constant of TiCl$_3$ in AAA, ABA, ABC is found to vary between 6.14 $\AA$ to 6.153 $\AA$,
while the inter-layer separation is found to vary between 5.850 $\AA$ to 5.866 $\AA$}. The weak van der Waals interaction of $\sim$ 1 meV between the layers allows the structure to transform from one stacking type to another, depending on the synthesis procedure. Since we are primarily interested in the in-plane physics of the hexagonal net of d$^{1}$ transition metals, for simplicity we will assume AAA stacking of the compounds in the rest of the discussion.

As per the available literature~\cite{cryst7050121} on the layered MX$_3$ compounds, these compounds have been reported to adopt either in the rhombohedral BiI$_3$ type~\cite{Devidas_2014}, or in monoclinic AlCl$_3$ structure type~\cite{ketelaar1947crystal}. In the BiI$_3$ structure, the honeycomb net is regular due to the three-fold symmetry, with three equal length metal-metal bonds. In the AlCl$_3$ structure, on the other hand, the honeycomb net can be distorted, and the y-coordinate of the M site determines the degree of distortion. This results in two unique in-plane M − M distances. Most compounds report a temperature driven transition from uniform rhombohedral BiI$_3$ type to bond-dimerized monoclinic AlCl$_3$ structure. The available data~\cite{troyanov1991x}   shows this structural transition typically happens around 100-200 K for d$^{1}$ transition metal tri-halides. In the discussion in the following, we will focus on the AAA stacked high temperature structure of uniform honeycomb net of transition metals possessing C$_3$ symmetry, which would host different possible electronic instabilities. As discussed later, subtle structures in the electron band structure paves the way to electronically driven charge-density-wave instabilities including dimerization, to which the lattice may react. 

Although considered structures hold a uniform hexagonal network of metal atoms, the underlying rhomohedral symmetry allows for the trigonal distortion of the MX$_6$ octahedra,  which occurs as the elongation or compression along one of the four three-fold symmetry axis, as shown in Figure \ref{fig:stacking}(b). The elongation (compression) results in decrease (increase) of the angle ($\theta$) between the M-X bond and the three-fold axis from ideal value of 54.73$^{o}$. Figure \ref{fig:stacking}(b) shows the crystal splitting of the $d$-levels of M atoms. In presence of the octahedral splitting, the five-fold degenerate $d$-levels split into three-fold degenerate $t_{2g}$ and two-fold degenerate $e_{g}^{\sigma}$. In presence of trigonal distortion of the MX$_6$ octahedra, the $t_{2g}$ levels further split into singly degenerate $a_{1g}$ and doubly degenerate $e_g^{\pi}$, with $\Delta_2$ and $\Delta_1$ denoting the octahedral and trigonal splitting, respectively.

\begin{table*}
\centering
\begin{tabular}{|p{2.2cm}|p{1.7cm}|p{1.5cm}|p{1.5cm}|p{1.5cm}|p{1.5cm}|p{1.5cm}|p{1.5cm}|p{1.5cm}|p{1.5cm}|}
 \hline
  &  \multicolumn{3}{|c|}{TiX$_3$} & \multicolumn{3}{|c|}{ZrX$_3$} & \multicolumn{3}{|c|}{HfX$_3$} \\
 \hline
  &F &Cl &Br  &F &Cl &Br  &F &Cl &Br\\
 \hline
 %\rowcolor{green}
 a=b(\AA)  & 5.202 & 6.030 & 6.405 & 5.487 & 6.204 & 6.524 & 5.320 & 6.096 & 6.423 \\
 \hline
 %\rowcolor{green}
  $\Delta{\theta}{_{trig}}$ (in degree) & 2.078 & -0.550 & -0.714 & 1.568 & -1.133 &  -1.563 & 0.815 & -1.244 & -1.563 \\
 \hline
 %\rowcolor{green}
  Exper. synthesis & No & Yes\cite{allegra1962calcolo} & Yes$^{*}$\cite{troyanov1990synthesis} & No & Yes$^{\dag}$\cite{swaroop1964crystal} & No & No & No & No \\
 \hline 
 \end{tabular}
\caption{In-plane lattice constants (a=b) and trigonal distortion
($\Delta{\theta}{_{trig}}$) of the theoretically optimized structures of MX$_3$. Given are also the information on experimentally synthesized compounds. $*$: structure with space group 148 (ABC stacking) is available, $\dag$:based on misaligned powder x-ray data.}
\label{table:lattice_parameters}
\end{table*}

As mentioned above, in the present study, we focus on the nine $d^1$ MX$_3$ compounds with M=Ti, Zr, Hf and X=F, Cl, Br, drawn from 3d, 4d and 5d transition metal series
and 2p, 3p and 4p series, as shown in Fig ~\ref{fig:stacking}(d).
The choice of these compounds, allows one to study the interplay of SOC and hopping, as well as Coulomb interaction. Moving
down the column, the SOC increases due to
increase in atomic number, as well as the metal-metal
hopping increases due to increase in the spatial extent of the
$d$-orbitals. Out of the nine proposed compounds, only three
compounds, namely TiCl$_3$~\cite{allegra1962calcolo,troyanov1991x,troyanov1991x}, TiBr$_3$~\cite{troyanov1990synthesis} and ZrCl$_3$~\cite{swaroop1964crystal} have been experimentally synthesized. To predict the crystal structure of the remaining compounds, we use the available experimental structure of the related compound with either common M or common X as the template and accordingly substitute the metal or the halide atom with the desired element. The constructed structure is subsequently fully relaxed by relaxing the atomic coordinates and the volume, fixing the symmetry. The influence of exchange-correlation and van der Waals interaction were checked in terms calculations within GGA, GGA+U and GGA+U+D2. Among these, GGA calculation consistently was found to reproduce well the experimentally measured in-plane lattice constants for compounds that have been synthesized. Since we are primarily interested in in-plane physics, in subsequent analysis we consider GGA optimized crystal structures in  P$\bar{3}$1m space group. The optimized values of in-plane lattice parameters as well as trigonal distortion($\Delta{\theta}{_{trig}}$) are given in Table ~\ref{table:lattice_parameters}. The in-plane lattice shows an expansion in moving from 2p to 3p to 4p, and from 3d to 4d to 5d. Interestingly, we note the trigonal distortion in fluoride compounds is opposite to that in chloride and bromide.
%%%%%%%%%%%%%%%%%%%%%%
\section{DFT Band structure}
\label{band-structure}

\begin{figure*}%[h!]
\centering
\includegraphics[scale=0.6]{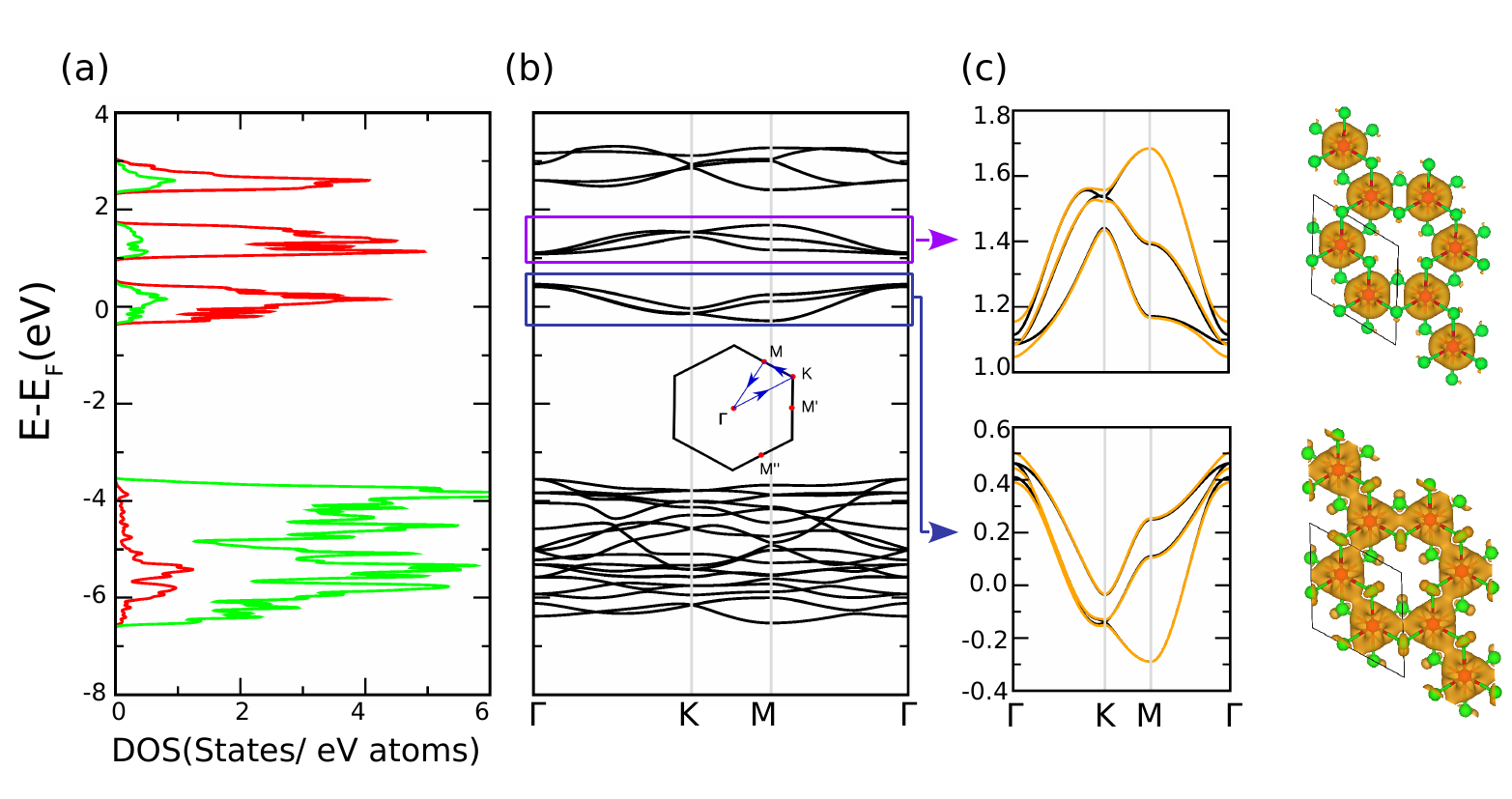}
\caption{\label{fig:frog} (a) GGA density of states of ZrCl$_3$, projected to  Zr-d (red) and Cl-p(green) states. (b) The corresponding band-structure plotted along the high symmetry points of the hexagonal BZ (shown as inset). (c) Zoomed-in band structure of bonding and antibonding $t_{2g}$ bands and the associated charge densities, with iso-surface value chosen to be 0.009 e$^{-}/{\AA}^3$. The zoomed-in band structure plot also includes the comparison of GGA (black) and GGA+SOC (yellow) bands. The zero of the energy in the density of states and band structure plots are set at corresponding Fermi energy.}
\label{zrcl3_band}
\end{figure*}

The GGA density of states of the representative compound ZrCl$_3$, over a broad energy scale of - 7 eV below Fermi level (E$_F$) and 4 eV above E$_F$, projected onto Zr $d$ and Cl $p$ states is shown in Fig.~ \ref{zrcl3_band}(a). We find a large separation between the predominantly X $p$-states (beyond $\sim 3.5$ eV below E$_F$) and predominantly M $d$ states (around E$_F$ and above), as expected for early transition metal halides. The corresponding band structure, plotted along the high symmetry points of the hexagonal Brillouin zone (BZ) (cf. inset), is shown in Fig.~\ref{zrcl3_band}(b). As evident from the plot, there are ten spin degenerate bands arising from the d-orbitals of the two M ions in the unit cell that clusters into three groups of $3,3$ and $4$ spin degenerate bands. While the highest energy group of four bands belong to $e_{g}^{\sigma}$ symmetry, the lower two groups of three bands (cf. Fig. \ref{zrcl3_band}(b,c)), spanning energy range of $\sim$ -0.3 eV to 0.5 eV, and $\sim$ 1 eV to 0.7 eV respectively, are of $t_{2g}$ symmetry. Since the trigonal crystal field splits three $t_{2g}$'s 
into 2+1 at each site, this cannot account for the above splitting of the six $t_{2g}$ bands  into 3+3. At any rate, the trigonal splitting arising from a distortion of 1-2$^\circ$ (cf Table. \ref{table:lattice_parameters}) is expected to much smaller compared to the splitting between the two groups, which is about 0.5 eV, as seen in the plot. To resolve this issue, we compute the charge densities corresponding to the 3+3 groups of bands which are shown in Fig.~\ref{zrcl3_band}(c). As is evident from the charge density plots, the grouping of six $t_{2g}$ bands into 3+3 arises due to bonding-antibonding combination of $t_{2g}$ orbitals resulting from highly directional, direct overlap of $t_{2g}$ orbitals along the three M-M bonds. The band structure plots in Fig.~\ref{zrcl3_band}(c), zoomed onto two $t_{2g}$ manifolds, show also the computed band structure including SOC. As is seen, SOC has a negligible effect on the entire band structure except lifting degeneracy at the high symmetry points, $\Gamma$ K, and M, which suggest strength of SOC to be relatively 
weak, and far from the assumption of infinite SOC limit. 

\begin{figure}
    \centering
    \includegraphics[width=1.0\columnwidth]{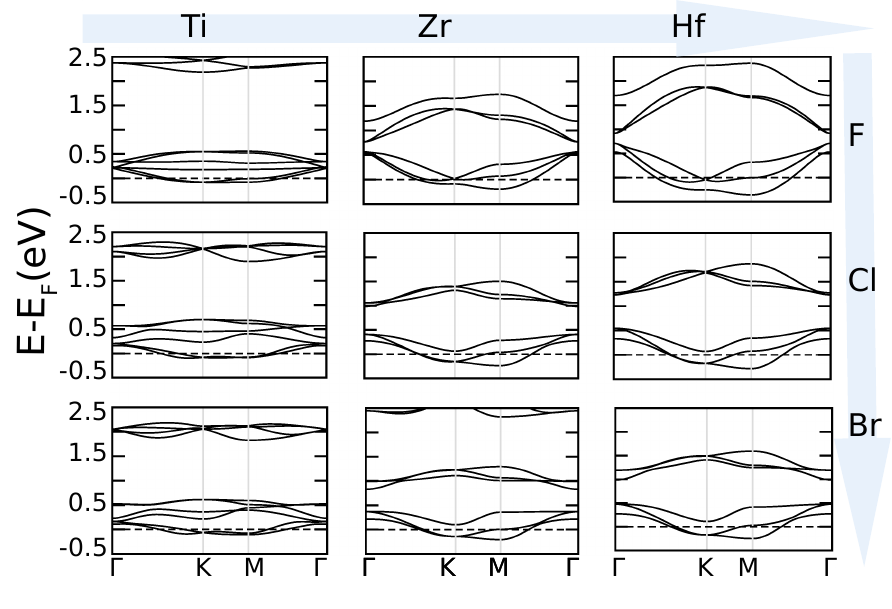}
    \caption{The variation in band-structure of MX$_3$, upon change of metal
    ion from 3d Ti $\rightarrow$ 4d Zr $\rightarrow$ 5d Hf, along the row,
    and change of halogen from 2p F $\rightarrow$ 3p Cl $\rightarrow$ 4p Br,
    calculated within GGA. Fermi energy is set to zero in the plots.}
\label{bands_lmto}
\end{figure}

The above discussed broad features of the electronic structure are found 
in other compounds as well, though they differ in details. For a comprehensive analysis of the $t_{2g}$ manifold of band structure of all nine MX$_3$ compounds, see Fig. \ref{bands_lmto}.  Moving along the metal series, we find while Zr and Hf share similar band structure features, that of Ti is different in terms of the individual bandwidths of the bonding and antibonding blocks, as well as in the separation
between bonding and antibonding blocks. This translates into weakening of the
direct metal-metal hopping and relative strengthening of indirect hopping via
halide ion in Ti compounds, compared to Zr or Hf. Moving along the halogen series, fluoride compounds exhibit markedly different dispersion compared to chloride or bromide counterparts, hinting electronic properties of fluorides to be different from chloride/bromide. A bulk of the discussion, therefore, will concentrate on the chlorides and bromides while we summarize briefly the fluorides in Sec. \ref{sec_flourides}.

\section{Effective Low energy tight binding model}
\label {TB} 

With the above DFT results, we now turn to the low energy modeling of the above band structure via effective low energy tight-binding Hamiltonian for the $t_{2g}$ orbitals in the presence of SOC.

According to the molecular orbital theory~\cite{Daintith_2008}, any two t$_{2g}$ orbitals on adjacent sites can interact to form six levels as ${\sigma}^{*}$, ${\pi}^{*}$, ${\delta}^{*}$, ${\sigma}$, ${\pi}$, ${\delta}$ where, we rank them from the highest to the lowest energy. The energy levels $\sigma$ and ${\sigma}^{*}$ are the consequence of the direct head-on overlap of the d-orbitals lobs called $dd\sigma$, whereas direct lateral overlap of d orbitals known as $dd\pi$ and $dd{\delta}$, gives rise to ${\pi}^{*}$, ${\delta}^{*}$, ${\pi}$, ${\delta}$ levels respectively. Among these, $\delta$ bonds are weakest and neglected henceforth.
These overlaps, along with indirect overlaps via the halide ion, dictate the nature of the resultant tight-binding model in $t_{2g}$ basis for single electron kinetic energy which is the starting point for our low energy analysis.

\begin{figure}
\centering
\includegraphics[width=0.8\columnwidth]{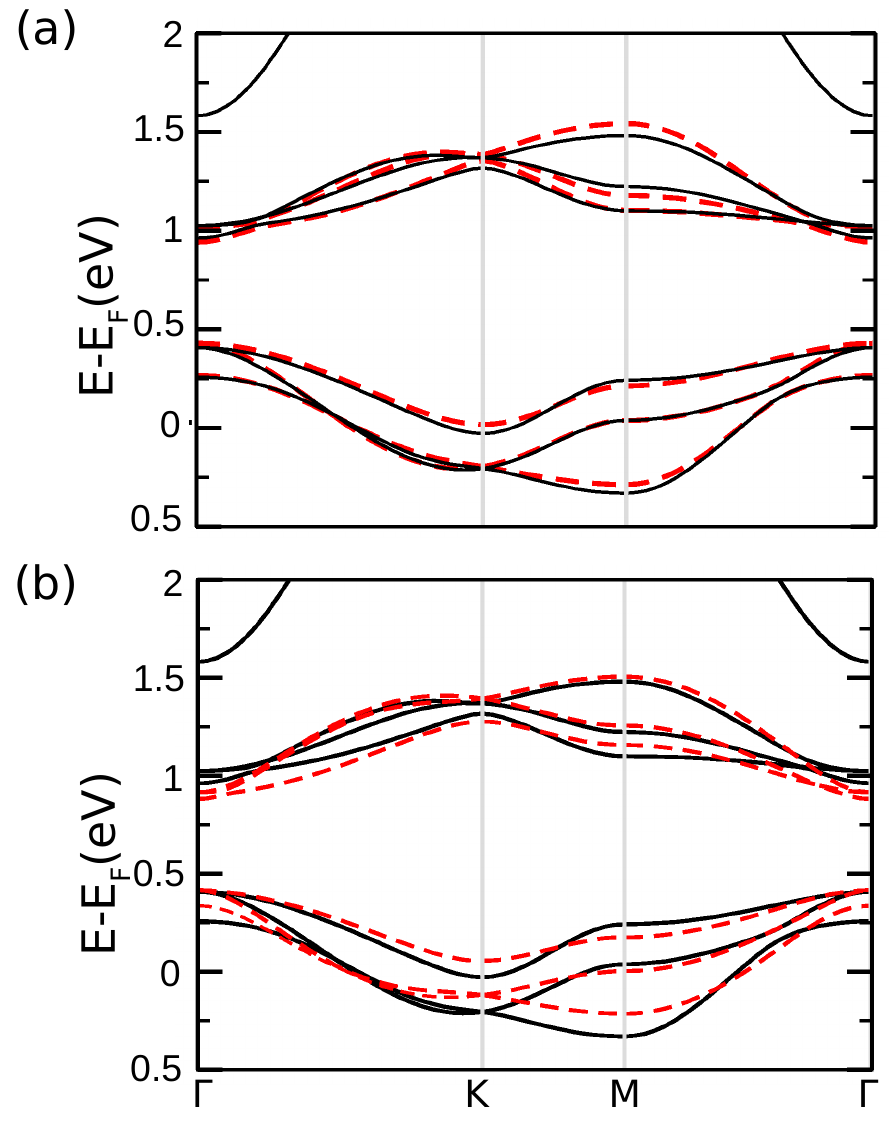}
\caption{(a) Downfolded $t_{2g}$ bands (in red) of ZrCl$_3$, obtained by integrating all of the degrees of freedom other than Zr $t_{2g}$ in comparison to the DFT band structure (in black). (b) The tight-binding bands (in red), obtained by restricting the real-space representation of the downfolded $t_{2g}$ Hamiltonian to only nearest neighbor Zr-Zr hopping, in comparison to the DFT band structure (in black). Zero of the energy is set at Fermi energy in (a) and (b).}
\label{fig:downfolded bands}
\end{figure}

In order to provide the realistic estimates of these direct and indirect overlap mediated hopping integrals, we derived the low-energy Hamiltonian in the effective transition metal $t_{2g}$ Wannier basis, starting from the full DFT band structure. For this purpose, we constructed the effective 
Wannier functions, by keeping only the metal $t_{2g}$ degrees of freedom in the basis and integrating
out the rest through the NMTO-downfolding technique. Fig. \ref{fig:downfolded bands} (a) shows the
comparison of the band structure in the downfolded basis, in comparison to the full band structure.
The good comparison, justifies the effectiveness of the prescription followed in deriving the low-energy
model.

The real-space representation of the downfolded bands shows non-zero hopping amplitudes up to the 4-th nearest neighbor (NN) among metal ions. However, a more amenable minimal hopping model with only first NN interaction is sufficient to reproduce most of the qualitative and quantitative features of the bands, as can be seen in Fig.(\ref{fig:downfolded bands}) for ZrCl$_{3}$. Similar conclusion holds also for other materials.

% New hopping pathways figures 
\begin{figure}
    \centering
    \includegraphics[width=1\columnwidth]{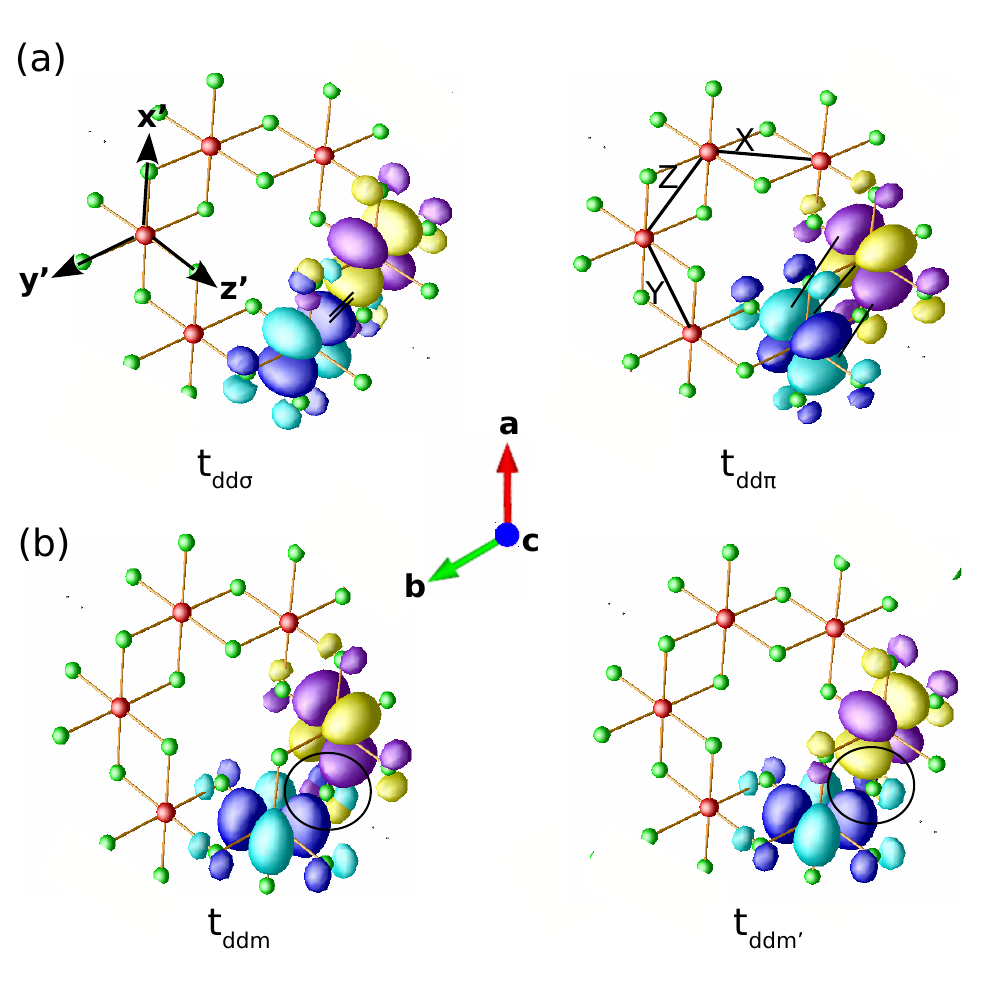}
    \caption{Overlap of effective $t_{2g}$ Wannier functions in the downfolded basis of ZrCl$_3$, placed at two neighboring positions of Zr ions. Lobes of opposite signs are colored as blue (yellow) and cyan (violet) for two different Zr ions. (a) Direct $dd$ overlaps, with left panel showing the head-on overlap ($dd\sigma$) and right panel showing the lateral overlap ($dd\pi$). The
    overlaps are highlighted by lines. The $t_{2g}$ orbitals are defined with the choice of local primed halogen-based coordinate systems, shown in left panel. The X, Y and Z M-M connecting bonds are shown in the right panel. The
    crystallographic coordinate system in terms of a, b and c is also shown.
    (b) Indirect overlaps via the halogen site, between the same (left) and
    differently shaped (right) $t_{2g}$ functions, labeled as  $ddm$ and $ddm'$
    in the text. Overlap region via halogen is encircled}
    \label{fig:hopping_pathways}
\end{figure}

The resultant first NN tight-binding Hamiltonian in $t_{2g}$ basis consists of four essential hopping amplitudes, as given below. For details of transformation between different coordinate systems see SM.\cite{SM} 
\begin{itemize}
\item The direct overlap between neighboring $t_{2g}$-orbitals that forms a $\sigma$-bond denoted by $t_{dd\sigma}$ (cf. 
Fig. \ref{fig:hopping_pathways}(a) left panel, shown for the representative case of ZrCl$_3$).
\item The direct overlap between neighboring $t_{2g}$-orbitals that forms a $\pi$-bond denoted by denoted by $t_{dd\pi}$ (cf. Fig. \ref{fig:hopping_pathways}(a) right panel, shown for ZrCl$_3$).
\item Two indirect overlap between neighboring $t_{2g}$-orbitals via the halides in the edge-sharing octahedral geometry denoted by $t_{ddm}$, $t_{ddm'}$, illustrated in  Fig.\ref{fig:hopping_pathways}(b), for ZrCl$_3$.
\end{itemize}

Note that in addition to the above hopping parameters, we also have the on-site trigonal energy scale. However, we find that in presence of dominant effect of $t_{dd\sigma}$ -- responsible for creating the bonding and the anti-bonding orbitals (cf. Fig.\ref{zrcl3_band}(c)), the effect of the trigonal distortion within each manifold of bonding or antibonding bands is minimal and hence we disregard it compared to the other hopping parameters. The results, discussed in the following, is not influenced by this assumption, as has been explicitly checked.

\begin{table*}%[h!]
\centering
\begin{tabular}{|p{2.2cm}|p{1.7cm}|p{1.5cm}|p{1.5cm}|p{1.5cm}|p{1.5cm}|p{1.5cm}|p{1.5cm}|p{1.5cm}|p{1.5cm}|}
 \hline
  &  \multicolumn{3}{|c|}{TiX$_3$} & \multicolumn{3}{|c|}{ZrX$_3$} & \multicolumn{3}{|c|}{HfX$_3$} \\
 \hline
  &F &Cl &Br  &F &Cl &Br  &F &Cl &Br\\
 \hline
$t_{dd\sigma}$ & -0.167 & -0.220 & -0.164 & -0.546 & -0.558 & -0.476 & -0.708 & -0.666 & -0.569 \\
 \hline
$t_{dd\pi}$ & 0.077 & 0.062 & 0.046 & 0.210 & 0.150 & 0.126 & 0.274 & 0.190 & 0.157 \\
\hline
$t_{ddm}$ & 0.058 & 0.078 & 0.079 & -0.061 & 0.030 & 0.038 & -0.126 & 0.015 & 0.031 \\
\hline
$t_{ddm'}$ & -0.022 & -0.030 & -0.027 & -0.023 & -0.020 & -0.020 & -0.039 & -0.022 & -0.020 \\
\hline
 $t_{dd\pi}$/$t_{dd\sigma}$ & -0.463 & -0.281 & -0.277 & -0.385 & -0.269 & -0.264 & -0.387 & -0.285 & -0.277 \\
 \hline
$t_{ddm}$/$t_{dd\sigma}$ & -0.351 & -0.357 & -0.481 & 0.112 & -0.053 & -0.081 & 0.178 & -0.022 & -0.054  \\
\hline
$t_{ddm'}$/$t_{dd\sigma}$& 0.134 & 0.136 & 0.163 & 0.043 & 0.037 & 0.042 & 0.055 & 0.033 &
       0.035 \\
       \hline 
 $\lambda$ & 0.015 & 0.028 & 0.040 & 0.030 & 0.043 & 0.043 & 0.060 & 0.152 & 0.152 \\
 \hline 
$\lambda/t_{dd\sigma}$   &-0.089 & -0.127 & -0.244 &-0.055 & -0.076 & -0.090 & -0.085 &
       -0.228 & -0.267 \\
 \hline 
$\Delta$=E$_{e_{g}}$-E$_{A_{1g}}$ &-0.030 & 0.050 & 0.039 & -0.178 & 0.057 & 0.077 & -0.212 & 0.065 & 0.088 \\
 \hline 
$\Delta/t_{dd\sigma}$&  0.180  & -0.227 &-0.236 &  0.325 &-0.103 & -0.161 &  0.300 &
       -0.097 & -0.154 \\
 \hline 
 \end{tabular}
\caption{DFT estimated hopping terms defined for $h_z$ matrix (cf. Eqn(2)) and SOC strength($\lambda$).
The last two rows show energy splitting in t$_{2g}$ level due to trigonal distortion. Apart from ratios, all the numbers quoted are in unit of eV.
}
\label{table:Mastertable}
\end{table*}

With this, one can now write down the first NN hopping model for the $t_{2g}$ orbitals as

\begin{align}\label{eq:tbd Hamiltonian without SOC}
H_{tb}=
\sum_{\langle ij\rangle} \sum_{\alpha,\beta} \sum_{\eta\eta'}~\Psi_{i\alpha\eta}^\dagger \left[h_{ij}^{\alpha\beta} \delta_{\eta \eta'}  \right]\Psi_{j\beta\eta'}.
\end{align}
Here $\Psi_{j\alpha\eta}$ annihilates electrons at the $j$'th site of the lattice with spin $\eta$ ($= \uparrow,\downarrow$), in the orbital $\alpha$ ($=xy,yz,zx$). The $h_{ij}$ is a $3\times 3$ Hermitian matrix at the bond connecting the $i$'th and the $j$'th sites of the lattice as shown in Fig.~\ref{fig:hopping_pathways}(a). 

Keeping in mind the different kinds of overlaps of the $t_{2g}$ orbitals, we can write the $h_{ij}$ matrix for the $Z$-bond as
\begin{eqnarray}\label{eq_general_hopping_t2g_zbond}
&h_Z &= \begin{pmatrix}
t_{dd\sigma}& t_{ddm'}& t_{ddm'}\\
t_{ddm'}& t_{dd\pi}& t_{ddm}\\
t_{ddm'}& t_{ddm}&	 t_{dd\pi}
\end{pmatrix} \nonumber\\
&& = t_{dd\sigma} h_\sigma + t_{dd\pi} h_{\pi} + t_{ddm}h_m + t_{ddm'}h_{m'}
\end{eqnarray}
where $t_{dd\sigma}, t_{dd\pi}, t_{ddm}, t_{ddm'}$ are hopping due to direct and indirect overlaps of the orbitals as discussed before. Also, $h_{\sigma},h_{\pi},h_{m}$ and $h_{m'}$ are $3\times 3$ matrices given by

\begin{eqnarray}
&&h_\sigma=\begin{pmatrix}
    1& 0& 0\\
    0& 0& 0\\
    0& 0& 0
\end{pmatrix},\quad
h_\pi= \begin{pmatrix}
    0& 0& 0\\
    0& 1& 0\\
    0& 0& 1
\end{pmatrix},\nonumber\\
&&h_m=\begin{pmatrix}
    0& 0& 0\\
    0& 0& 1\\
    0& 1& 0
\end{pmatrix},\quad
h_{m}'=\begin{pmatrix}
    0& 1& 1\\
    1& 0& 0\\
    1& 0& 0
\end{pmatrix}.
\end{eqnarray}
The form of the hopping of the X and Y bonds can be obtained by exploiting the three-fold rotation symmetry of the lattice, as detailed in the SM~\cite{SM}.

The DFT estimates for the hopping amplitudes, $t_{dd\sigma}$ etc. for different materials are given in Table~\ref{table:Mastertable}. For different materials, the generic hierarchy of the relative strengths of the hopping parameters are found to be as follows 
\begin{eqnarray}
|t_{dd\sigma}| > |t_{dd\pi}| \gg |t_{ddm}| \geq |t_{ddm'}|.
\end{eqnarray}

The low energy effective tight binding model can be obtained by adding atomic SOC to Eq. \ref{eq:tbd Hamiltonian without SOC}, which, after appropriate scaling, we write as :
\begin{align}\label{eq:minimal model for phase diagram}
\mathcal{H}=\mcl{E}\sum_{\langle ij\rangle} \sum_{\alpha \beta}\sum_{\eta\eta'}~\Psi^\dagger_{i\alpha\eta}\left[H_{ij}^{\alpha\beta}\delta_{\eta\eta'}-\tilde{\lambda} {\bf l}_{\alpha\beta}\cdot{\bf s}_{\eta\eta'}\delta_{ij}\right]\Psi_{j\beta\eta'}.
\end{align}
where we have introduced the overall energy-scale 
\begin{eqnarray}
    \mcl{E} = |t_{dd\sigma}| + | t_{ddm}|.
\end{eqnarray}
such that
$H_{ij}$ can be obtained by re-scaling $h_{ij}$ (Eq.\ref{eq:tbd Hamiltonian without SOC}) and in particular its form on the $Z$ bond is obtained by re-scaling Eq. \ref{eq_general_hopping_t2g_zbond} as 
\begin{align}\label{eq_minimal_ham_2}
H_{ Z}=-(1-\tau_m)h_\sigma + \rho\tau_m h_m + r (1-\tau_m) h_\pi + \tau_m' h_{m'}
\end{align}
with
\begin{align}
&t_{dd\sigma}=-\mcl{E} (1-\tau_m);~~~~~~t_{dd\pi}=r ~t_{dd\sigma}; ~~~~t_{ddm}=\rho\mcl{E}\tau_m;\nonumber \\
&t_{ddm'} = \mcl{E} \tau_m';~~~~\lambda=\mcl{E}\tilde{\lambda}.
\label{eq_tbparam}
\end{align}
$\lambda (>0)$ is the strength of SOC and $\mbf{l}$'s represent the three $l=1$ orbital angular momentum matrices while $\mbf{s}$ are the Pauli matrices that represent the spin degrees of freedom of the electrons. To estimate the SOC in the studied materials, we calculated the band structure within GGA+SOC. The tight-binding fit of the obtained band structure with DFT-derived hopping integrals together with a tunable $\lambda$ was used to extract the best fit $\lambda$ value of a given compound. The estimated $\lambda$ values are listed in Table.\ref{table:Mastertable}.

The Hamiltonian in Eq. \ref{eq:minimal model for phase diagram} leads to a rich set of possibilities even at the non-interacting level which crucially decides the fate of electron-electron interactions and the low energy phases. While in actual materials all the parameters are present, it is instructive to investigate the above tight binding model in steps by gradually incorporating different parameters starting with $\tau_m$ and $\tilde{\lambda}$, as taken in the following sections and subsections. 

To this end, we note that the parameter $\rho=\pm 1$ indicates that the indirect hopping amplitude $t_{ddm}$ can be of either sign. In particular we find that (see Table \ref{table:Mastertable}) ZrF$_3$ and HfF$_3$ have $\rho=-1$, making the situation markedly different
from that of chlorides/bromides, which was already hinted from the
band structure (see Fig. 3). This distinct difference of the fluorides arises from the following characteristic features of F -- (1) much smaller ionic radius (147 pm), and (2) much higher electronegativity (3.98); compared to chlorine and bromine with ionic radii (electronegativities) of 175 pm (3.16) and 185 pm (2.96) respectively.
Since none of the fluoride compounds have been so far synthesized, 
for the rest of this paper, we shall be concerned with $\rho=+1$ and shall take up the case for fluorides towards the end in Sec.~\ref{sec_flourides}.

\section{Single electron phase diagram}
\label{phase-diagram}

Our first principle calculations show that the major deviation from the indirect hopping model discussed in Ref. \cite{mondal2023emergent,PhysRevLett.121.097201} is the direct hopping given by $t_{dd\sigma}$. Hence it is useful to gain further insights into the interplay of the direct hopping, indirect hopping and the SOC. This is obtained by using $r=\tau'_m=0$ in Eq. \ref{eq_minimal_ham_2} such that $\tau_m=0(1)$ corresponds to the purely direct (indirect) hopping limits at different values of SOC. The importance of direct metal-metal interaction, over the conventional description of ligand-mediated interaction 
in description of phenomenology of transition metal compounds, has been
acknowledged in recent time, in context of cobaltates.~\cite{PhysRevB.106.165131} The resultant phase diagram forms the basis to understand the material relevant final phase diagram which we obtain by sequentially turning on $t_{dd\pi}$ and $t_{ddm'}$.

\subsection{Phase diagram for \texorpdfstring{$\tau_m-\tilde\lambda$}{} model}

\begin{figure*}%[h!]
%    \centering
    \includegraphics[scale=.59]{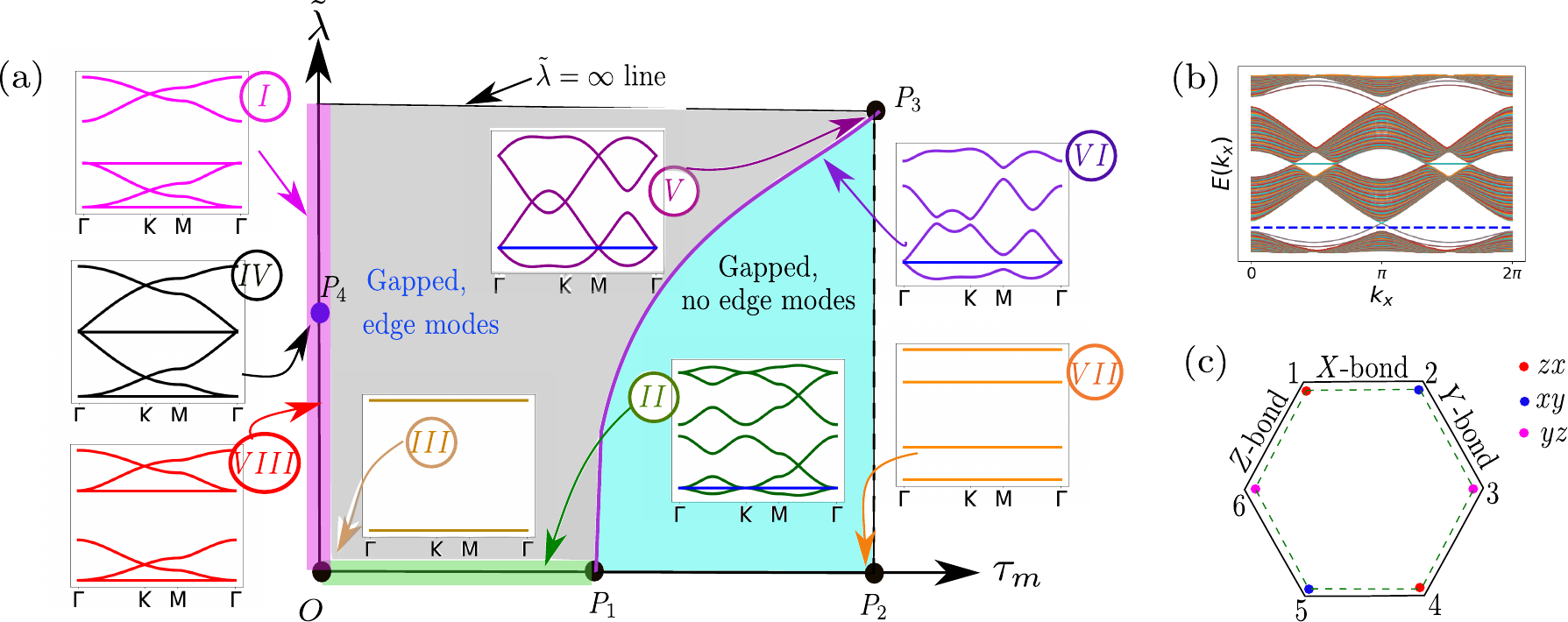}
    \caption{ (a) Phase diagram in $\tau_m-\tilde{\lambda}$ plane for  $r=0,\tau_m'=0$, $\rho=+1$. The two gapped phases are shown with different colors (gray and cyan). The band structures, plotted along the $\Gamma$, K and the M points of the BZ. at different points on the phase diagram are shown as insets. The blue horizontal lines in the insets~$II$, $V$ and $VI$ show the position of the Fermi level. In the inset~$II$, both the bands are six-fold degenerate as explained in the text. The insets~$V$ and $VI$ are for very large value of $\tilde \lambda$ and hence the only the four lower energy $J=3/2$ bands are shown. (b) Band structure in the topological gapped phase (for $\tau_m=0.9, \tilde\lambda = \infty, r=0, \tau_m'=0$) in cylindrical geometry. The edge modes at the Fermi-energy are shown by the blue solid line, with Fermi level marked in dashed line. (c) A single hexagon showing the origin of molecular orbitals at point $P_2$ of the phase diagram in Fig~\ref{fig_phase_diagram1}(a). Different colored dots represent three $t_{2g}$ orbitals. The six sites of the hexagon are labeled with numbers from 1 to 6. The symmetric linear combination of the orbitals connected by the green dotted line form the lowest energy band. Other orbitals are localized on singles bonds of the hexagon are shown by black dotted lines.
    See text for details.}
    \label{fig_phase_diagram1}
\end{figure*}

The phase diagram in the $\tau_m-\tilde{\lambda}$ plane is shown in Fig.~\ref{fig_phase_diagram1}(a). The top right corner, $P_3 (\equiv (\tau_m=1,\tilde\lambda=\infty$) corresponds to infinite SOC in the purely indirect hopping limit with infinite coupling. This, for $d^1$ gives rise to SU(8) Dirac semi-metal (DSM) as discussed in Ref. \cite{mondal2023emergent}. At $P_3$, the six $t_{2g}$ orbitals (including spin degeneracy) split up into four $J=3/2$ and two $J=1/2$ orbitals which are separated by infinite energy gap ($\propto\tilde\lambda$) with the $J=3/2$ orbitals being of lower energy. Hence at this point we obtain 1/4th filled $J=3/2$ orbitals whose band structure is shown in inset $(V)$ of Fig. \ref{fig_phase_diagram1}(a). Here, the lowest band linearly touches the upper band at the $\Gamma$ and the M points of the BZ giving rise to four 4-component Dirac fermions sitting at four valleys -- the three M points of the BZ and one at the $\Gamma$ point that constitutes the SU(8) DSM~\cite{mondal2023emergent}.  Remarkably, almost the entire phase diagram, except the pink and green shaded parts along the $\tau_m=0$ and $\tilde\lambda=0$ axis, can be understood from this SU(8) limit as we now discuss.

On moving away from the SU(8) point, all the Dirac fermions get gapped out, generically giving rise to band insulators. However the nature of these two band insulators obtained in the two extreme limits of changing $\tau_m$ or $\tilde\lambda$ away from $P_3$, are different. One of them -- that obtained by varying only $\tau_m$ -- is a free fermion symmetry protected topological phase (SPT) \cite{Senthil_SPT}, as is evident from the gapless edge modes plotted in Fig.~\ref{fig_phase_diagram1}(b). These edge modes are protected by time reversal symmetry. Indeed out of the 24 distinct ways of gapping out the SU(8) Dirac fermions discussed in Ref. \cite{mondal2023emergent}, there are precisely two different time reversal invariant lattice  ($\mathcal{A}_{1g}^e$) singlet masses {where we have used the notations of Ref. \cite{mondal2023emergent} for ready reference}. The above two band insulators correspond to these two singlets as detailed in SM~\cite{SM}. The two insulating phases are separated by a phase transition denoted by the magenta curve connecting the points $P_3$ and $P_1\equiv (\tau_m=0.67,\tilde\lambda=0)$. On this line, only the Dirac fermions at the $\Gamma$ point becomes gapless while those at the three $M$-points remain gapped across the transition as shown in the band-structure $(VI)$ in Fig. \ref{fig_phase_diagram1}(a). The resultant theory has an enhanced SU(2) symmetry as detailed in SM~\cite{SM}. 

In Fig. \ref{fig_phase_diagram1}(a), diametrically opposite to $P_3$ is the point $O\equiv (\tau_m=0,\tilde\lambda=0)$ which describes the purely direct hopping model via $t_{dd\sigma}$ without SOC. Here, the Hopping Hamiltonian (Eq. \ref{eq:minimal model for phase diagram}) reduces to a particularly simple form, given by 
\begin{eqnarray}
&&H = -\mathcal{E} \left(\sum_{\Braket{ij}\in \text{Z-bonds}}  \Psi^\dagger_{i,xy,\eta} \Psi_{j,xy,\eta} + \right. \nonumber \\
&&\left.\sum_{\Braket{ij}\in \text{X-bonds}}  \Psi^\dagger_{i,yz,\eta} \Psi_{j,yz,\eta} + \sum_{\Braket{ij}\in \text{Y-bonds}}  \Psi^\dagger_{i,zx,\eta} \Psi_{j,zx,\eta}\right) \nonumber \\
&&\qquad ~+~ h.c.\nonumber\\
\end{eqnarray}
such that on the $Z/X/Y$-bonds (See Fig.~\ref{fig:hopping_pathways}(a)) respectively only the $xy/yz/zx$-orbitals hop. Since each set of bonds forms a disconnected network of dimers that rotate into itself under $C_3$, we get bonding and anti-bonding orbitals of the respective types on each of the three bonds resulting in two sets of 6-fold (including spin $\eta=\uparrow,\downarrow$) degenerate flat band as shown in inset $(III)$ of Fig. \ref{fig_phase_diagram1}(a). This is evident in the form of the Wannier functions obtained from DFT (cf. Fig. \ref{fig:hopping_pathways}). This kind of separation of the energy bands into two groups of three bands is also seen in the DFT band structures of ZrCl$_3$ shown in Fig.~\ref{zrcl3_band}(c), which is dominated by the direct overlap $t_{dd\sigma}$ as given in Table~\ref{table:Mastertable}. 

The fact that the entire $OP_1$ segment on the $\tilde\lambda=0$ line is gapless is expected on very general grounds and is in fact dictated by the general structure of the phase diagram starting from the SU(8) symmetric point, $P_3$. This can be rationalized based on that fact on this line there is an enhanced SU(2) spin rotation symmetry such that this line cannot be a part of the free fermion SPT lying above it for finite $\tilde\lambda$-- as predicted by the SU(8) theory. The trivial insulating phase (shown in cyan in Fig. \ref{fig_phase_diagram1}(a)) of course can be connected to the spin-rotation symmetric segment $P_1P_2$ continuously. Finally at the point $P_2\equiv (\tau_m=1,\tilde\lambda=0)$, the bands become flat again with a degeneracy of 2-4-4-2 (inset $(VII)$ of Fig. \ref{fig_phase_diagram1}(a)). At this point ($P_2$), the lowest band is made up of spin-degenerate {\it molecular orbitals} of the type shown in Fig. \ref{fig_phase_diagram1}(c) at each hexagon. On deviating from this point, these orbitals acquire dispersion. Hence the entire gapped trivial insulator (shown in cyan in Fig. \ref{fig_phase_diagram1}(a)) can be understood in terms of these effective eigenmodes. 

\begin{figure*}%[h!]
	\centering
    \includegraphics[scale=0.6]{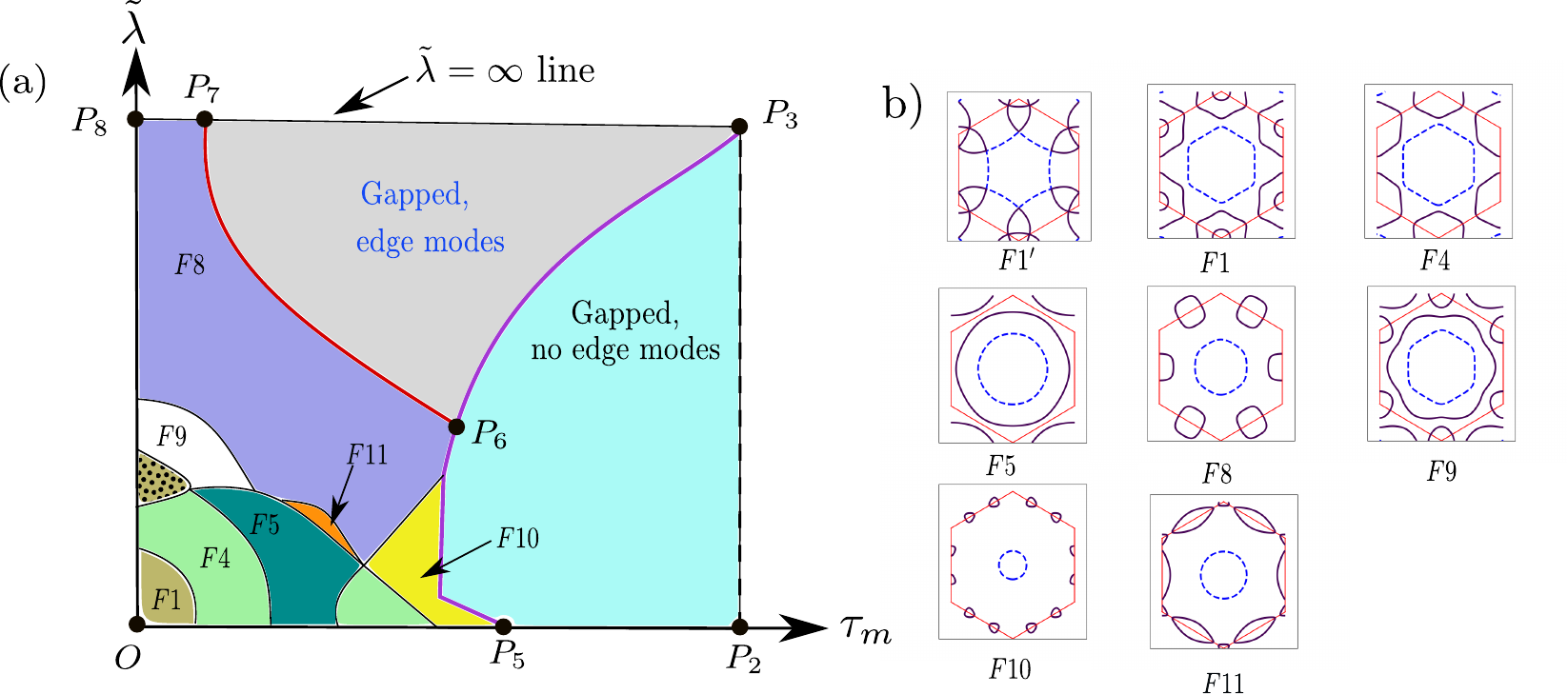}
	\caption{(a) Phase diagram for $\rho=+1 ,r = 0.3, \tau_m'=0$. Two gapped
 phases shown in gray and cyan shading. The different metallic phases are shown with different colors and labeled as $F1$, $\ldots$, $F11$.
 (b) Fermi surfaces
 corresponding to different metallic phases. The hole-like Fermi surfaces are shown with blue dashed lines and the electron-like FS are shown with solid violet lines. The hexagonal BZ is shown with red solid lines. The $F1^{'}$ Fermi surface corresponds to the $O$ point of the phase diagram shown in panel (a). The un-dotted and dotted region of $F1$
 differ by the fact for the Fermi surface corresponding to the dotted  region, one of the Fermi pockets around the K points is electron-like and the other is hole-like, while both are electron-like for the un-dotted region.}
 \label{fig_phase_diagram_2}
\end{figure*}

On increasing the SOC ($\tilde\lambda$) along the $\tau_m=0$ line in Fig. \ref{fig_phase_diagram1}(a), the six-fold symmetry is independently lifted in the bonding and anti-bonding sectors without intermixing for small $\tilde\lambda$ as shown in inset $(VIII)$ of the figure. The band structure (inset $(VIII)$) is very similar to that of monolayer Kagome band structure~\cite{guterding2016prospect} -- for both the bonding and anti-bonding sectors-- with the lower dispersing band touching the flat band quadratically at the $\Gamma$-point of the BZ. As one increases the SOC, the band-width of each of the two sectors increases while retaining their overall shape such that at the point $P_4\equiv(\tau_m=0,\tilde\lambda=1.35)$ the bands touch at the $\Gamma$ point leading to a spin-1 Dirac dispersion~\cite{PhysRevA.83.063601} at the at the touching of the two sectors (inset $(IV)$ of Fig. \ref{fig_phase_diagram1}(a)). On increasing SOC further, remarkably the second flat band -- previously associated with the anti-bonding sector -- detaches from it and becomes a part of the bonding sector leading to a division of four lower bands and two higher bands (inset $(I)$ of Fig. \ref{fig_phase_diagram1}(a))-- as expected from the $J=1/2$ and $J=3/2$ splitting of the atomic orbitals at large SOC. For $d^1$ materials, however, the above change of band structure is not important as only the lowest flat band is filled such that the chemical potential lie at the lowest quadratic band touching points leading to a very unstable (to interactions) quadratic band-touching semimetal with one of the flat bands having divergent effective mass.

The above structure of the phase diagram gives a good starting point to connect to the DFT band structure by incorporating the sub-leading interactions as we now turn to discuss. Two such important sub-leading parameters are $t_{dd\pi}$ and $t_{ddm'}$ representing the sub-leading direct and indirect hopping respectively (see Eqs. \ref{eq_general_hopping_t2g_zbond} and \ref{eq_tbparam} as well as Table. \ref{table:Mastertable}). We study their effects {as a build up to the material phase diagram}.

\subsubsection{Effect of \texorpdfstring{$t_{dd\pi}$}{}} 

The first sub-leading hopping that is relevant  across all the compounds is the direct hopping via the $\pi$-overlap denoted by $t_{dd\pi}$ as shown in Fig. \ref{fig:hopping_pathways} and incorporated via the parameter $r=|t_{dd\pi}|/|t_{dd\sigma}|$ in our effective tight-binding Hamiltonian (Eq. \ref{eq:minimal model for phase diagram}) as shown in Eq. \ref{eq_tbparam}. However, instead of scanning the entire phase diagram as a function of $r$, we shall confine ourselves to $r=0.3$-- a value which is roughly consistent for the different materials. The resultant phase diagram is shown in Fig. \ref{fig_phase_diagram_2}.

The $P_2P_3$ line of Fig. \ref{fig_phase_diagram_2}(a) is exactly equivalent to that of Fig. \ref{fig_phase_diagram1}(a) and hence the description of the entire trivial gapped band insulator in cyan region remains same apart from the quantitative renormalization of the band structure away from the $\tau_m=1$ line. Similarly the physics of the $\tilde\lambda=\infty$ for $\tau_m<1$ holds until the point $P_7$ giving rise to the $Z_2$ free fermion SPT (gray region) with gapless edge modes, exactly in the case of $t_{dd\pi}=0$ in Fig. \ref{fig_phase_diagram1}(a). The intermediate line, $P_3P_6$, hence is associated with a Dirac band-touching at the $\Gamma$-point of the BZ giving rise to a SU(2) DSM. However, the effect of $t_{dd\pi}=r|t_{dd\sigma}|\propto (1-\tau_m)$ drastically rearranges the band structure for lower $\tau_m$, as we discuss now.

The $t_{dd\pi}$ lifts the threefold degeneracy of the flat bands at the point $O=(\tau_m=0,\tilde\lambda=0)$ leading to dispersive bands that cross the chemical potential giving rise to a compensated band metal such that the net Luttinger volume is zero. The relevant Fermi surface is named $F1'$ and is shown in Fig.~\ref{fig_phase_diagram_2}(b). However this is highly unstable due to the touching of the hole and particle Fermi pockets and on increasing both $\tau_m$ and $\tilde\lambda$, the resultant Fermi surface undergoes topological changes  giving rise to a plethora of compensated band metals denoted by $F1-F11$ in Fig. \ref{fig_phase_diagram_2}(a). The intervening Lifshitz transitions~\cite{lifshitz1960anomalies,volovik2017topological} include cases where both separate sheets of Fermi surfaces merge, e.g. $F4$ to $F5$ via van-Hove singular necks, as well as, instances where individual sheets of Fermi surfaces disappear,
e.g. $F1$ to $F4$. This generic appearance of the compensated band metals with diverse Fermi-surface topology is particularly relevant to the materials under consideration as we discuss in the next section in detail along with the relevant Lifshitz transitions.  

We would like to end this discussion about the effect of $t_{dd\pi}$ by commenting on the metals $F1-F11$ (Fig. \ref{fig_phase_diagram_2}(a)) that occupy the region that was erstwhile (Fig. \ref{fig_phase_diagram1}(a)) a part of the topological insulator.  Interestingly for $F8$, the  electron bands evolve continuously from the free fermion SPT and hence it inherits a non-trivial $Z_2$ invariant for the bands crossing the chemical potential. In fact, except on the $\tilde\lambda=0$ line, we find that for all the metals in the phase diagram under consideration, one of the bands crossing the chemical potential has non-trivial $Z_2$ index, calculated following the method discussed in Ref.~\cite{fuKane_Z2number}. The method is applicable for systems with inversion symmetry, as is in the present case.

\subsubsection{Effect of \texorpdfstring{$ t_{ddm'}$}{}}

\begin{figure}
    %\centering
    \includegraphics[scale=0.475]{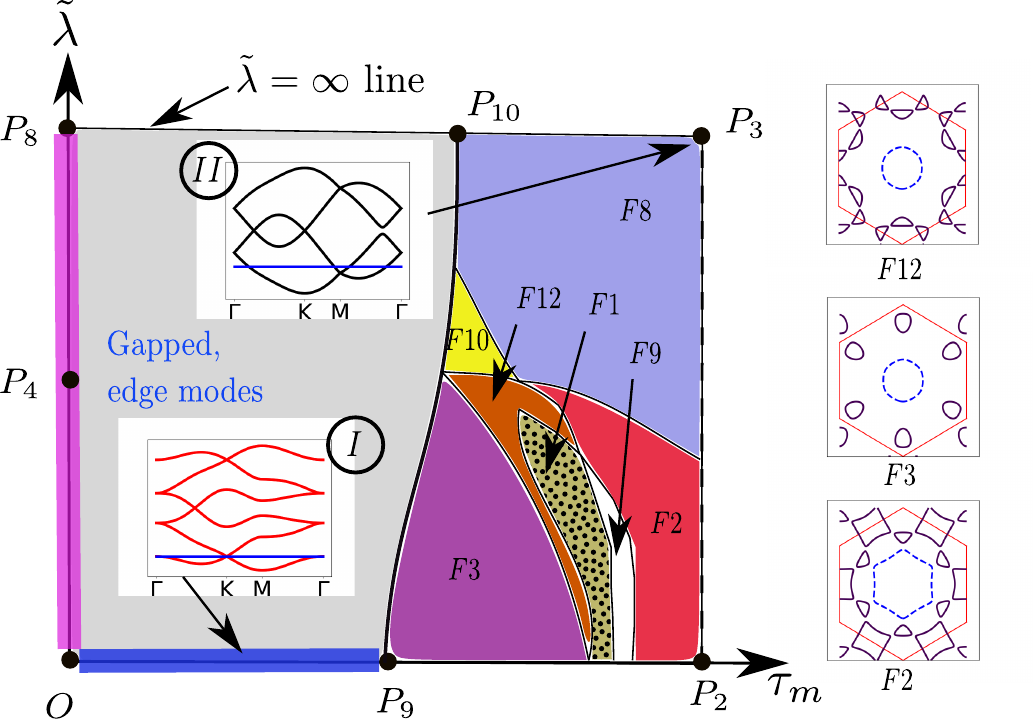}
    \caption{Phase diagram for $r=0,\tau_m'=-\tau_m$, $\rho=+1$. Inset $I$ shows a prototypical band structure for a point on the $OP_9$ line where the lowest band touches the next band linearly at the K points, while the inset $II$ shows the band structure at the $P_3$ point. Since $\tilde\lambda = \infty$ at $P_3$, only four bands ($J=3/2$ bands) are shown in inset~$II$. As in Fig.\ref{fig_phase_diagram_2}(a), metallic phases with different Fermi surface
    topology, are labeled and colored. In addition to $F8$, $F9$ and $F10$,
    introduced in Fig.\ref{fig_phase_diagram_2}(a), three new phases, labeled
    as $F2, F3$ and $F12$ appear. The Fermi surface topology for $F2, F3$ and $F12$ phases are shown by the side. }
    \label{fig:phase diagram 3}
\end{figure}

We now turn to the effect of the indirect hopping mediated by $ t_{ddm'}$ on the minimal model with phase diagram in Fig. \ref{fig_phase_diagram1}(a). Again we choose a representative value of $t_{ddm'}=-t_{ddm}$-- in the regime relevant to the materials --  to indicate its effect. Unlike $t_{dd\pi}$, this indirect hopping now drastically reorganizes the $\tau_m\approx 1$ region of the phase diagram-- apparent by contrasting Fig. \ref{fig_phase_diagram1}(a) with Figs. \ref{fig_phase_diagram_2}(a) and \ref{fig:phase diagram 3}.

In particular the line $\tau_m=0$ remains unaltered with respect to the minimal model (Fig. \ref{fig_phase_diagram1}). Also the free fermion SPT (in gray in Fig. \ref{fig:phase diagram 3}) is stable to finite $t_{ddm'}$, albeit it does not extend all the way to the point $P_3$. In fact the SU(8) Dirac point ($P_3$) now develops into a compensated band metal as the Dirac cones, at the erstwhile $P_3$ point, moves away from the chemical potential in opposite direction-- the Dirac node at the $\Gamma$ point moves above it and those at the $M$ points move below it -- giving rise to Fermi pockets around these points. The band structure at the $P_3$ point is shown in inset~$II$ of the phase diagram in Fig.~\ref{fig:phase diagram 3}. The resultant Fermi surface around this point is of $F8$ type given in Fig.~\ref{fig_phase_diagram_2}(b).

 As we move away from the $P_3$ point along the $\tilde \lambda=\infty$ line by decreasing $\tau_m$ (thus increasing $t_{dd\sigma}$), the F8 Fermi surface continues to exist, although a finite gap at the M and the $\Gamma$ opens up (away from the chemical potential) between the lowest and the second lowest band. This gap opening due to the effect of $t_{dd\sigma}$ gives a non-trivial $Z_2$ number to the lowest band and hence the $F8$ metal in the phase diagram in Fig.~\ref{fig:phase diagram 3} is a topological metal. On moving further away from the $P_3$ point along the $\tilde \lambda=\infty$ line, the size of the Fermi pockets of the $F8$ metal continuously shrink and eventually vanish at the $P_{10}$ point after which the system enters into a gapped phase. Since this transition from the $F8$ metal to the gapped phase does not happen through a band touching, the $Z_2$ invariant of the lower band remains unchanged across this transition and hence the gapped phase is also topological insulator-- the same free fermion SPT as in Fig. \ref{fig_phase_diagram1}.

 Turning to low SOC, the $P_2$ point no longer has flat bands but now gains a dispersion due to $\tau'_m$ leading to a compensated metal of Fermi surface type $F2$ as shown in Fig. \ref{fig:phase diagram 3}. On moving from the point away from the $P_2$ point along the $\tau_m=1$ line, the $F2$ Fermi surface transforms into a $F8$ type Fermi surface, which is then connected to the $P_3$ point. This $F8-F2$ transition does not involve any band touching, but just a change of the chemical potential and hence the $F2$ region of the phase diagram is also a topological metal. On reducing the values of $\tau_m$ from the point $P_2$, the system encounters various other metallic phases which have different Fermi surfaces ($F9,F1,F12,F3$ etc.). We find that for all these phases, the bands crossing the Fermi energy always have a nontrivial $Z_2$ index. Thus, all the metals in this phase diagram are also topological metals.

 Finally, on the line $OP_8$, $\tau_m = \tau_m' = 0$ and hence the description of this line is the same as in the phase diagram in Fig.~\ref{fig_phase_diagram1}. On the other hand, along the $OP_9$ line (for which $\tilde \lambda=0$), the lowest band touches the upper band linearly at the K points. The effect of finite $\tilde \lambda$ opens up gap at the K points and the system enters into the topological  gapped (gray shaded region).

\subsection{The material phase diagram}

Having discussed the minimal tight-binding model and the effect of the sub-leading hopping terms resulting in a rich single-particle phase diagram, we now turn to the regime that may be most suited to the material parameters, except for the fluorides. To this end we choose the representative hyperplane given by $ \rho=+1, r=0.3, \tau_m' = -\tau_m$ and vary $\tau_m\in (0,1)$ and $\tilde\lambda \in (0,\infty)$.

\begin{figure}
    \centering
    \includegraphics[scale=0.55]{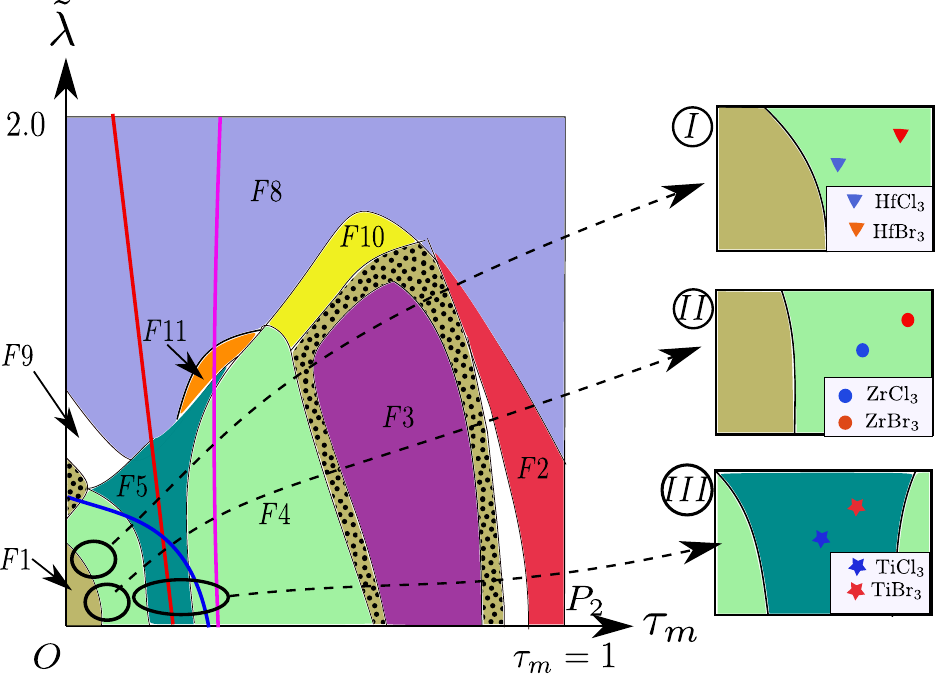}
    \caption{Phase diagram for $\rho=+1, r=0.3, \tau_m' = -\tau_m$ with $0 \leq\tau_m \leq 1$ and $0\leq \tilde \lambda \leq 2.0$. For $\tilde \lambda>2.0$, the $F8$ phase continues to exist. The six different chloride/bromide compounds are placed in this phase diagram, according to
    the estimated parameter values of the low energy Hamiltonian (cf. Table~\ref{table:Mastertable}), shown in the insets $I,II$ and $III$.
    The phase diagram is demarcated by the red, blue and magenta lines, according
    to the $Z_2$ characters of the bands. See text for details.}
    \label{fig:final phase diagram}
\end{figure}

 The phase diagram in this parameter regime is shown in Fig.~\ref{fig:final phase diagram}. Due to the complementary effects of the secondary direct and indirect hopping -- $t_{dd\pi}$ and $t_{ddm'}$ respectively -- the resultant phase diagram is in a way
 superposition of Figs. \ref{fig:phase diagram 3} and \ref{fig_phase_diagram_2}(a) such that all the phases appearing in this case are gapless, perfectly compensated and have Fermi surfaces with at least one partially filled band having non-trivial $Z_2$ invariant.

Based on which particular band(s) carry non-trivial $Z_2$, the phase diagram can be demarcated by red, magenta and blue lines (see Fig.~\ref{fig:final phase diagram}). The $Z_2$ index for the lowest band is non-zero for the region of the phase diagram which is in the right hand side of the red solid line. On the other hand, the second lowest band has nontrivial $Z_2$ index for the regions of the phase diagram which are either left to the red solid line or right to the magenta solid line. On the red line, the lowest and the second lowest bands touch at the M point and the $Z_2$ character of the two bands switch. On the magenta line, the second lowest band touches the third lowest band and thus encounters another change in $Z_2$ character. The third lowest band, which crosses the Fermi energy only at the $F1$ region which is near the origin $O$, has non-trivial $Z_2$ index for the region which is left to the blue solid line in the phase diagram. On this blue line, the third lowest band touches the fourth lowest band and encounters a change in $Z_2$ character.
 
 The positioning of the materials ZrX$_3$, TiX$_3$ and the HfX$_3$ (X = Cl, Br) in the phase diagram, based on the estimated parameters given in Table~\ref{table:Mastertable} is shown
 in zoomed plots given in Fig.~\ref{fig:final phase diagram}. Due
 to weaker SOC compared to the strength of leading hopping interactions, the studied compounds are all placed towards the
 bottom of the phase diagram. Given the fact, that $t_{dd\sigma}$
($t_{ddm}$) is significantly larger(smaller) in Zr/Hf compounds compared
to Ti compounds, as expected, Zr and Hf compounds are placed left to
Ti compounds. Given the similarity in electronic structure of
Zr and Hf compounds (see Fig .3), it is not surprising that they belong 
to the same $F4$ class,
with Hf compounds lying higher in position compared to Zr, due to
stronger SOC. On the other hand, Ti compounds belong to distinctly
different $F5$ class. Systematically, bromine compounds lie higher and
right to chlorine compounds, due to stronger SOC and weaker direct
hopping strength, respectively.
 
\begin{figure}
    \centering
    \includegraphics[scale=0.3]{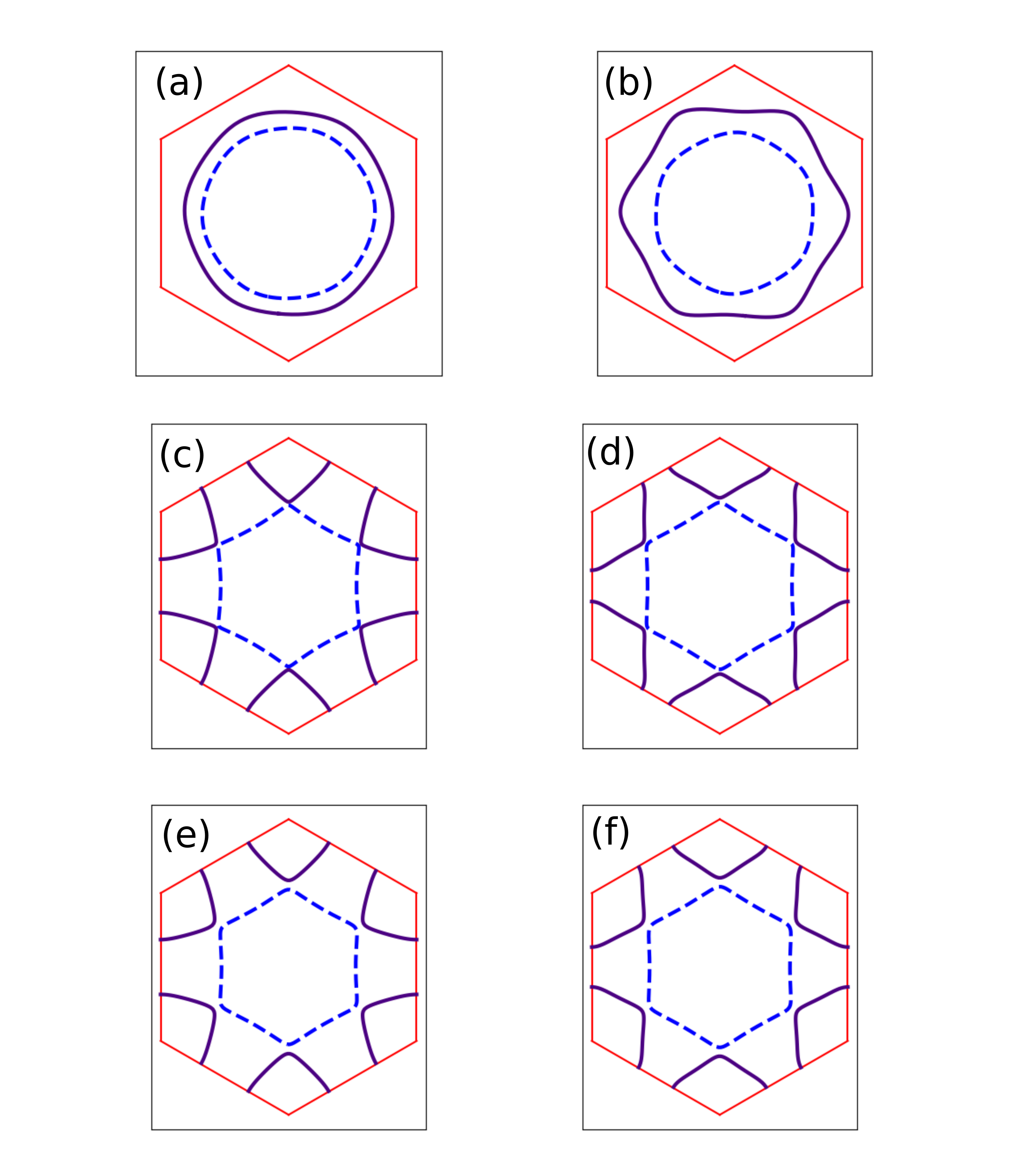}
    \caption{Fermi surfaces for (a)~TiCl$_3$, (b)~TiBr$_3$, (c)~ZrCl$_3$, (d)~ZrBr$_3$, (e)~HfCl$_3$, (f)~HfBr$_3$. The (a) and (b) are $F5$-type Fermi surface while the rest are $F4$ type. The blue dotted lines show hole-like Fermi surface and the solid indigo colored lines show electron-like Fermi surface. The red line shows the boundary of the hexagonal BZ.}
    \label{fig:material_FS}
\end{figure}

The detailed Fermi surface(FS)s of the compensated, topological metallic phases of the six compounds are shown in Fig.~\ref{fig:material_FS}.  The $F4$ type FS of ZrX$_3$ and HfX$_3$ compounds, is characterized by three disjoint Fermi pockets-- two electron-like pocket around the two K points and one hole-like pocket around the $\Gamma$ point. On the other hand, the TiX$_3$ compounds having $F5$-type FS, have two Fermi pockets, one electron-like and one hole-like around the $\Gamma$ point.
 
\subsection{Nesting and Lifshitz transitions}

A characteristic feature of some of the FSs in Fig. \ref{fig:material_FS} are the flattish {\it almost nested} sections -- involving both intra and inter-pockets. This makes them particularly susceptible to nesting instabilities in presence of electron-electron interactions at appropriate wave-vectors. Our preliminary results indeed indicate enhanced susceptibilities in the charge-density-wave channel due to such nesting. The detailed characterization of such instabilities though require more accurate study particularly due to the intricate structure of the FS's involved.

Another feature of the phase diagram is the presence of plethora of Lifshitz phase transitions~\cite{lifshitz1960anomalies,volovik2017topological} between the variety of compensated metals  (Fig. \ref{fig:final phase diagram}). These phase transitions involving a change in the Fermi surface topology can be classified into two broad categories~\cite{lifshitz1960anomalies, yamaji2006quantum} -- (1) {\it pocket vanishing} type associated with disappearance of new segments of Fermi surface {\it e.g.}, between $F1$ and $F4$ where the Fermi pockets centered around the BZ corners appear-- possibly relevant for Zr(Hf)Cl$_3$ and Zr(Hf)Br$_3$, and, (2) {\it neck collapsing} type associated with merging of two segments of Fermi surfaces {\it e.g.}, the transition between $F4$ and $F5$ where two particle-like Fermi pockets develop a neck that meets at the M points possibly relevant for TiCl$_3$ and TiBr$_3$. These transitions, accessed, in the present case, by tuning the band parameters at a particular filling, occur due to the change of the band-structure at the chemical potential. While the former leads to a step function in the single-particle density of states, $\rho(\epsilon-\epsilon_F)\sim \theta(\epsilon-\epsilon_F)$, the latter has a logarithmic singularity, {\it i.e.}, $\rho(\epsilon-\epsilon_F)\sim -\ln |\epsilon-\epsilon_F|$ and hence has a van Hove singularity arising from the vanishing Fermi velocity for the electrons on the Fermi surface.
This singular behavior can be reflected in thermodynamic measurements such as magnetic susceptibility~\cite{yamaji2006quantum} as well as scaling of bipartite entanglement entropy~\cite{PhysRevB.87.115132}. Interestingly, the tuning of the band parameters can be achieved through bi-axial straining which should be achievable considering the layered structure of the materials similar to SrRuO$_4$\cite{sunko2019direct}. Considering about
2$\%$ compressive strain on ZrCl$_3$, the direct $dd\sigma$ hopping
is found to enhance by about 20$\%$ while the indirect hopping is found to be heavily suppressed, thereby conducive to triggering a $F4$ $\rightarrow$ $F1$ transition. This may be even easier for Hf compounds, which are even closer to the $F4$-$F1$ boundary. Our DFT calculated FS for 1$\%$ strained Hf compound, indeed shows a $F1$ type.
Straining on Ti compounds shows similar effect, although the percentage change is found to be much smaller. Therefore such straining may be of interest in investigating the physics of the Lifshitz transition.

\section{\texorpdfstring{$\rho$ = -1}{} : Implication for Fluorides}
\label{sec_flourides}

\begin{figure}%[h!]
    \centering
    \includegraphics[scale=0.5]{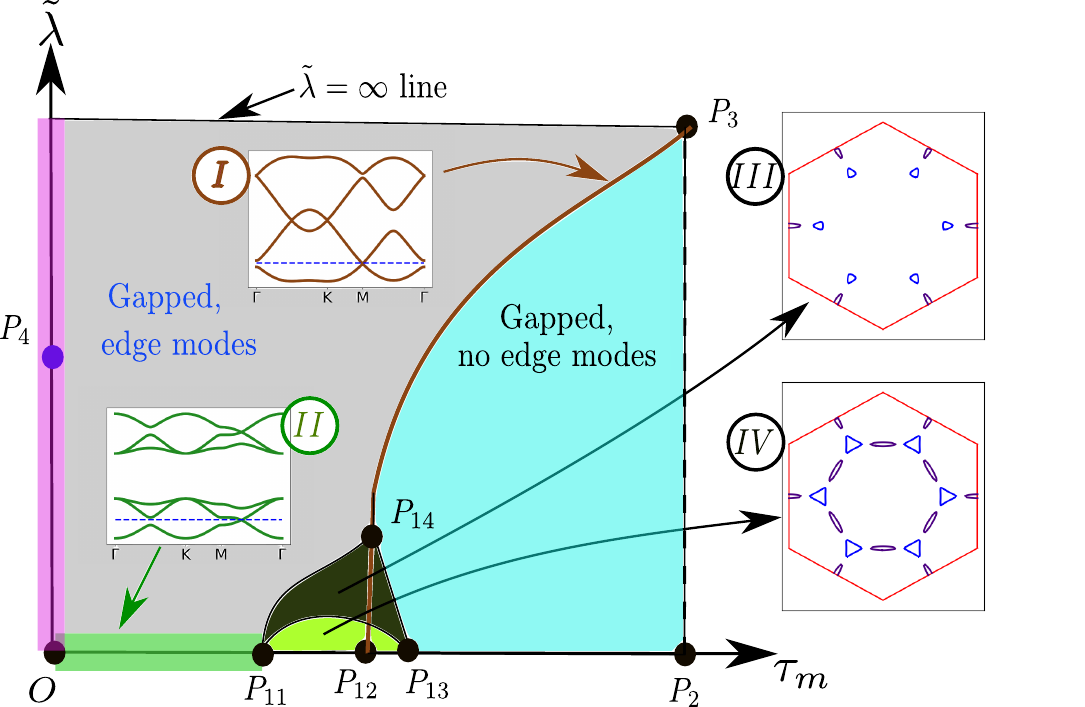}
    \caption{$\tau_m-\tilde{\lambda}$ phase diagram with $\rho=-1, r=0, \tau_m'=0$. The two gapped phases, shaded as gray and cyan,  as well as pink coloured lines in the phase diagram are identical to those
    in Fig~\ref{fig_phase_diagram1}(a). The band structures on a typical point on the $OP_{11}$ line is shown in inset~$II$. The inset $I$ shows the band structure on the transition line between the two gapped phases. Note that inset~$I$ shows only four bands since it is drawn for a large value of $\tilde \lambda$ where the four lower energy $J=3/2$ orbitals are separated from the higher energy $J=1/2$ orbitals by a large energy separation. As opposed to
    Fig~\ref{fig_phase_diagram1}(a), this phase diagram hosts
    two topological metallic phases, marked in lime-green and dark green shades. The Fermi surfaces for the two metallic regions are shown by insets $III$ and $IV$.}
    \label{fig:fluride_phase_diagram}
\end{figure}

Having discussed the situation with the chlorides and the bromides, we now turn to fluorides, which as indicated above (cf Table \ref{table:Mastertable}) show markedly different electronic structure. Furthermore, unlike, chlorides and bromides, the tight-binding parameters for fluorides show diverse behavior even 
among the 3d, 4d and 5d transition metals, the parameters for Ti being rather different from that of Zr/Hf. This hinders providing a universal
framework to describe the three fluoride compounds, captured 
through a common phase diagram, as was possible for chlorides and bromides. We thus concentrate on the most striking difference between Zr/Hf chlorides and bromides, and Zr/Hf fluorides, namely the change in sign of the indirect hopping,
$t_{ddm}$, captured by the parameter $\rho$ in Eq. \ref{eq_tbparam}. 
This affects some of the basic conclusions stemming from the structure of the minimal phase diagram, presented in Fig. \ref{fig_phase_diagram1}. In the following, we thus confine ourselves to the $\tau_m-\tilde{\lambda}$ phase diagram,  which determines the nature of the low energy single-particle starting point for these materials,
without delving into the complexity of the sub-leading hopping like $t_{dd\pi}$ and $t_{ddm}$.
The obtained results are shown in Fig. \ref{fig:fluride_phase_diagram}, which should be contrasted with Fig. \ref{fig_phase_diagram1}(a). 

First of all, we notice a similarity of the phase diagrams in Fig. \ref{fig:fluride_phase_diagram} and Fig. \ref{fig_phase_diagram1}(a), especially for large $\tilde\lambda$. This apparent similarity, however
hides an important contrast, that can be best understood as follows.
Starting from the SU(8) limit, $P_3$ in the present case, is a {\it particle-hole} inverted version of Fig. \ref{fig_phase_diagram1}(a) due to the change in sign of $t_{ddm}$. Thus, while the $P_3$ still gives a SU(8) DSM with four 4-component Dirac points at $\Gamma$ and  three M points, the associated spinors are not necessarily the same as in the previous case, but are related to it via a microscopic particle-hole transformation. In fact this theory is therefore a particle-hole conjugate version of the SU(8) metal discussed in Ref. \cite{mondal2023emergent} and hence the same mass analysis can be applied to the present case. It is for this reason, we still have the same two gapped phases-- the free fermion topological and trivial  band insulators on deviating away from the $SU(8)$ semi-metal-- resulting from the two lattice singlet masses~\cite{mondal2023emergent}.  However, the difference in the eigen modes due to the change in sign becomes apparent at the transition between the two insulators, which now is brought about by closing of the gap at the three M points on the line $P_3P_{14}$ while that at the $\Gamma$ point remains gapped-- unlike in case for $P_3P_1$ line in Fig. \ref{fig_phase_diagram1}(a). This leads to an enlarged SU(6) internal symmetry for the $N_F=3$ free Dirac fermions on the $P_3P_{14}$ line (see SM).

Another consequence of the above change in the nature of the spinors is the appearance of additional compensated semi metals for lower values of $\tilde\lambda$ around $t_{dd\sigma}\sim t_{ddm}$. The corresponding Fermi surfaces are shown in insets $III$ and $IV$. The lowest bands of these metals have non-zero $Z_2$ index on the closed region $P_{11}P_{12}P_{14}$, while the upper band is $Z_2$-trivial. On the line $P_{12}P_{14}$, the lowest band touches the upper band at the M points and the $Z_2$ indices of these two bands switch. Thus, one of the bands of these metals always has non-zero $Z_2$ index and these are topological metals. On the $P_{13}P_{14}$ line, Fermi pockets of the metals shrink to zero and the system enters the trivial insulating phase. Hence, the change of sign of the indirect hopping, makes the situation more favorable to the topological metallic phase, which may be stabilized even without inclusion of sub-leading hopping terms.
 
\section{Summary and Outlook}
\label{sec_summary}

In summary, following the stimulating proposal of achieving SU(8) DSM state in quarter-filled J=3/2 SOC coupled electrons, we critically examine the material realization of this proposal considering d$^{1}$ transition metal tri-halide family of compounds, MX$_3$. Systematic variation of both the metal site from 3d to 5d (M= Ti, Zr, Hf) as well as the halide site from 2p to 4p (X=F, Cl, Br) allows to study the interplay of different microscopic energy scales. Although only three out of the studied nine compounds have been synthesized so far, we do hope our study will encourage synthesis of other compounds too.

Our first-principle electronic structure calculations in combination with minimal tight-binding models show that a hierarchy of electron hopping pathways is needed to faithfully capture the rich low energy single electron physics in these compounds. Importantly, our study uncovers the dominant role of direct metal-metal hopping. Thus, a minimum of five  band parameters -- atomic SOC ($\lambda$), two direct metal-metal hopping ($t_{dd\sigma}, t_{dd\pi}$), two indirect metal-halide-metal hopping ($t_{ddm},t_{ddm'}$)-- dictate the low energy fermiology. Our study further unravels while the chlorides and bromides have a generic trend of band parameters,  the fluorides are distinct -- due to the drastic difference in the size and electronegativity of the fluoride ion compared to chloride and bromide.

Inclusion of this material specific reality renders the physics of the above candidate compounds in a domain far removed from the idealized SU(8) DSM. Instead, a variety of topologically 
non-trivial compensated metals gets stabilized upon variation of relative strength of direct vs indirect hopping and SOC, which differ in their 
Fermi surface topology. Fermi surfaces with different topology are found to be connected through intervening Lifshitz phase transitions.
Remarkably, though, the ideal SU(8) point serves as a useful starting point to understand the global structure of the above phase diagram. In particular, the compensated metallic phases are found to be asymptotically
connected to topological insulating states resulting from
gapping out of SU(8) semi-metals. Placing of the compounds, in the 
emergent in the $t_{dd\sigma}$/$t_{ddm}$ − $\lambda$ phase diagram reveals
the chloride compounds are close to the phase boundary separating two metals with different Fermi surface topology. Introduction of 
bi-axial strain in these layered compounds, is found to cause a large
variation in the $t_{dd\sigma}$/$t_{ddm}$, thereby could be an effective tool to induce Lifshitz transition in chloride compounds, in particular.

While the above study captures the physics of non-interacting electrons
in the un-dimerized lattice, several possible ordering instabilities can be triggered
in these compensated metallic phases at lower temperatures, as discussed in the following.
For instance, electron-electron
interaction driven ordering instabilities are expected to be particularly enhanced near the Lifshitz transitions~\cite{PhysRevB.68.245109,PhysRevB.82.045110,yamaji2006quantum} due to the singular nature of single-particle density of states. Akin to the neck collapsing Lifshitz transition and associated singular nature of the density of states found in present case, similar
phenomena have been discussed in Sr$_2$RuO$_4$, which reports enhancement of instability near Lifshitz point\cite{sunko2019direct}. 
The renormalization group calculations for the short-ranged four fermion interactions near the neck collapsing between two particle-like Fermi pockets indicate an enhancement of the BCS superconducting instability~\cite{PhysRevB.92.085112,PhysRevB.98.035122}.  In particular, the neck collapsing transition at BZ boundary at M points between $F4$ and $F5$ opens up the possibility of finite momentum instabilities in both the particle-hole and particle-particle channels. The former can lead to a charge-density wave insulator generically accompanied by dimerization as observed in some of the candidate materials~\cite{troyanov1991x, gapontsev2021dimerization}. The finite momentum pairing instability leading to a (Fulde-Ferrel-Larkin-Ovchinikov) FFLO-like \cite{PhysRev.135.A550,larkin1964nonuniform} phases  are equally interesting. These issues call for further detailed investigation.

Furthermore, the effect of short-ranged Hubbard interaction, $U$, supplemented with a Hund's scale, $J$, relevant for multi-orbital systems energetically favors, within our GGA+U+SOC calculations (see SM for details), a ferromagnetic metal in all the materials for lower values of $U-J$, which in turns gives way to ferromagnetic insulator and/or stripy and zig-zag spin density wave insulators for larger $U-J$ values.
This hints towards a profusion of competing interaction effects likely to be in play in the current family of materials, driving the electrons to various ordered states at lower temperatures. We further note the metallic state with net moment and topological character should lead to intrinsic anomalous Hall conductivity.

Finally, while the present study focuses on  d$^1$ honeycomb compounds, it is straightforward to extend these ideas to materials with d$^3$ configuration such as MoCl$_3$.  Our minimal tight-binding model suggests a similarly rich fermiology in such materials including a curious exchange of flat band along $\tau_m=0$ line in Fig. \ref{fig_phase_diagram1}. This will be taken up in future.

%%%%%%%%%%%%%%%%%%%%%%

\acknowledgements

  The authors acknowledge discussion with Arun Paramekanti, H. R. Krishnamurthy and Vijay Shenoy.  The authors acknowledge DST, Government of India (Nano mission) for funding under project no. DST/NM/TUE/QM-10/2019 (C)/7. MG acknowledges CSIR, India, for the senior research fellowship [grant no. 09/575 (0131) 2020-EMR-I]. BM and SB acknowledge Max Planck Partner group Grant at ICTS, Swarna Jayanti fellowship grant of SERB-DST (India) Grant No. SB/SJF/2021-22/12 and the Department of Atomic Energy, Government of India under Project No. RTI4001. SB acknowledges adjunct fellow program at SNBNCBS, Kolkata for hospitality.  T.S.-D. acknowledges JC Bose National Fellowship (Grant No. JCB/2020/000004) for funding.

\bibliography{ref}

\end{document}

\title{Supplemental Material:
Ab-initio Insights on the Fermiology of $d^1$ Transition metals in Honeycomb lattice : Hierarchy of hopping pathways and spin-orbit coupling}

\maketitle

\section{Transformation of Basis: Global to Halide-based, primed Coordinate System}

\begin{figure}
        \centering
        \includegraphics[width=0.5\columnwidth]{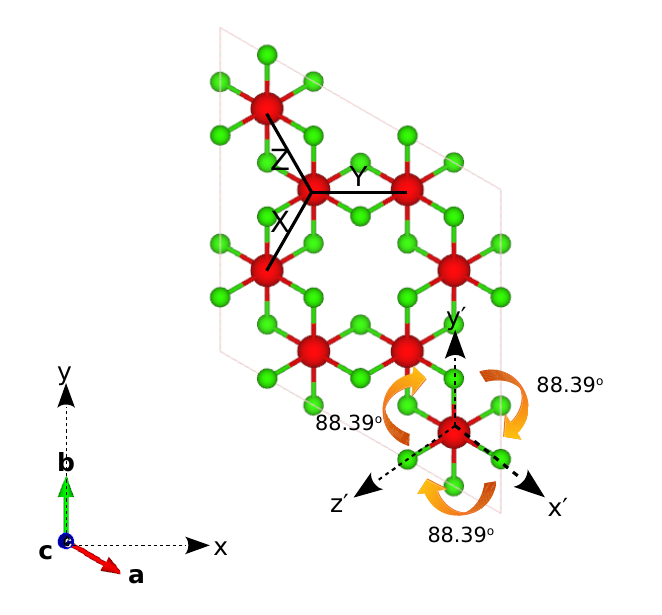}
        \caption{Global and local coordinate systems denoted by {$x$, $y$, $z$}, and {$x'$,$y'$,$z'$}, respectively. Red and green balls represent to metal (M) and halide (X) ions, respectively. A metal ion is connected to its nearest neigbours via X,Y, and Z bonds, also marked in the figure.}
        \label{fig:local_cordinate}
    \end{figure}

In the global coordinate system, the x-axis  points in the direction making 30$^{\circ}$ with \textbf{a}, y pointing along \textbf{b}, and z along \textbf{c}. A Local octahedral coordinate system is defined by $x'$, $y'$, and $z'$ axis, where the corresponding $x$-axis, $y$-axis, and $z$-axis point along the bonds M-X as shown in the Fig.(\ref{fig:local_cordinate}).
 
The transformation matrix rotating the global coordinate system to the local coordinate system, for ZrCl$_3$, is given by 
    $$x’=(-0.7089) x + (0.4051) y + (-0.5772) z$$
    $$y’=(0.0034) x + (-0.8165) y + (-0.5772) z$$
    $$z’=(-0.7052) x + (-0.4112) y + (0.5774) z$$
    
It is to be noted that due to trigonal distortion, the M-X bonds are not orthogonal to each other in the octahedron but have a mutual angle of 88.39$^{\circ}$. Thus the above transformation matrix aligns the $x'$, 
$y'$, and $z'$ axis, approximately, along the M-X bonds shown in the Fig.(\ref{fig:local_cordinate}). The rotation of the coordinate system, given above, gets 
reflected on d-orbitals via the transformation matrix as \\
    
    $d_{x’y’} =(0.5802) d_{xy}+ (0.2374) d_{yz} +(0.5772) d_{z2} +(0.4072) d_{xz} + (0.3283) d_{x2-y2}$\\
    
    $d_{y’z’}=(0.5744) d_{xy} + (-0.2340) d_{yz} +(-0.5774) d_{z2} +(0.4091) d_{xz} + (-0.3382) d_{x2-y2}$\\
    
    $d_{z’2}=(0.5023 ) d_{xy} + (-0.4113) d_{yz} +(0.0001) d_{z2} +(-0.7053) d_{xz} + ( 0.2842) d_{x2-y2}$\\
    
    $d_{x’z’}=(0.0058) d_{xy}+ (0.4713) d_{yz} +(-0.5774) d_{z2} +( -0.0022) d_{xz} + (0.6666) d_{x2-y2}$\\
    
    $d_{x’2-y’2}=(-0.2843) d_{xy} + (-0.7052) d_{yz} +(-0.0000) d_{z2} +(0.4112) d_{xz} + (0.5025) d_{x2-y2}$

%%%%%%%%%%%%%%%%%%%%%%%%%%    

    %%%%%%%%%%%%

\section{Derivation of $h_X, h_Y$ from $h_Z$}
Since the Hamiltonian, given in Eq.~1 of the main text, is symmetric under the action of $C_3$ rotation, the hopping matrices on the $Y$ and the $X$ M-M bonds can be obtained 
from that on the $Z$-bonds, by the transformation given by,
\begin{eqnarray}
    &&h_X = \mcl{R}_3^\dagger h_Z \mcl{R}_3 \\
    &&h_Y = \mcl{R}_3^\dagger h_X \mcl{R}_3. 
\end{eqnarray}

Here, $\mcl{R}_3$ is a $3\times 3$ unitary operator that implements $C_3$ rotation about an axis perpendicular to the plane of the honeycomb lattice, on the $t_{2g}$ orbitals and is given by
\begin{eqnarray}
    \mcl{R}_3 = \begin{pmatrix}
        0 & 1& 0\\ 0& 0& 1\\ 1& 0& 0
    \end{pmatrix}.
\end{eqnarray}
%%%%%%%%%%%%%%%%%%%%%%%%%%%%%

\section{Evolution of the Band structures in $\tau_m$-$\tilde \lambda$ Phase Diagram}

\subsubsection{The $\tilde\lambda=0$ line} 
\begin{figure}[h!]
	\includegraphics[scale=0.4]{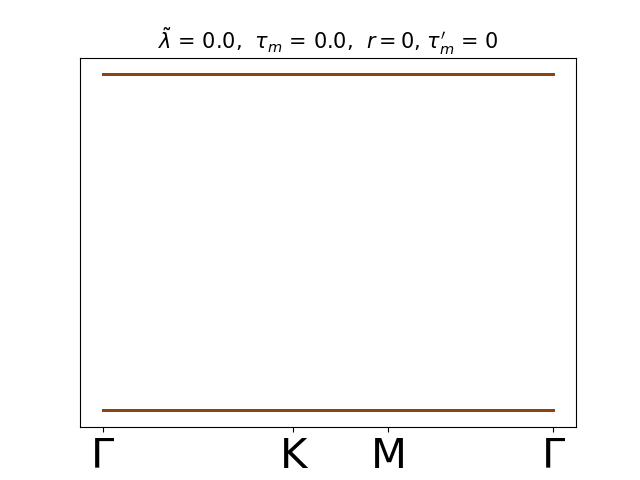} \qquad
	\includegraphics[scale=0.4]{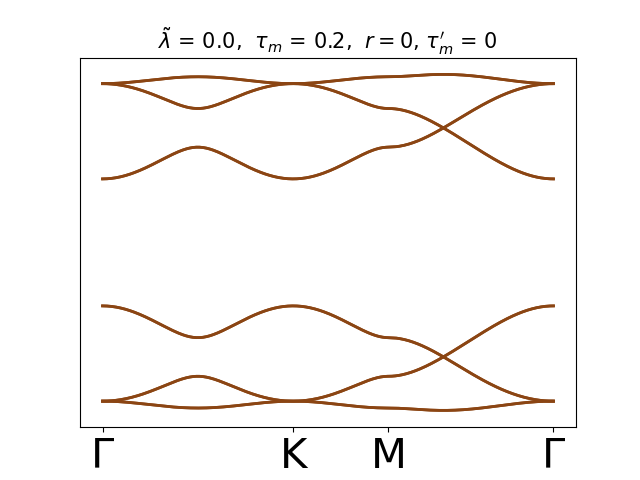} \\ \quad \\
	\includegraphics[scale=0.4]{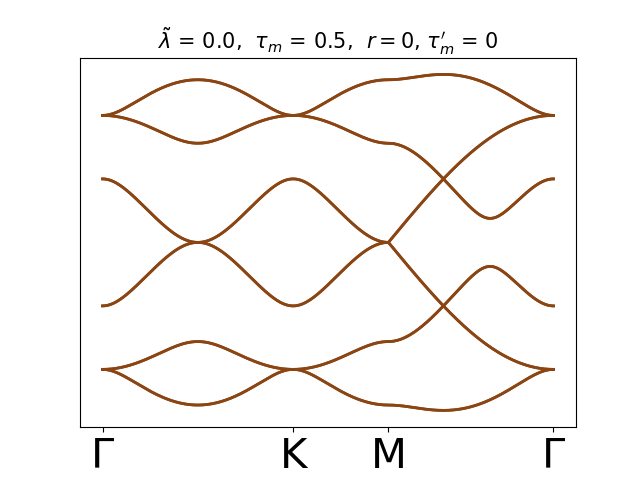} \qquad 
	\includegraphics[scale=0.4]{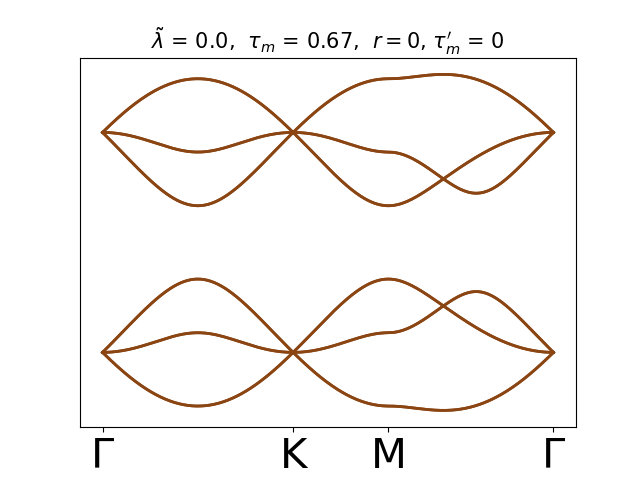} \\ \quad \\
	\includegraphics[scale=0.4]{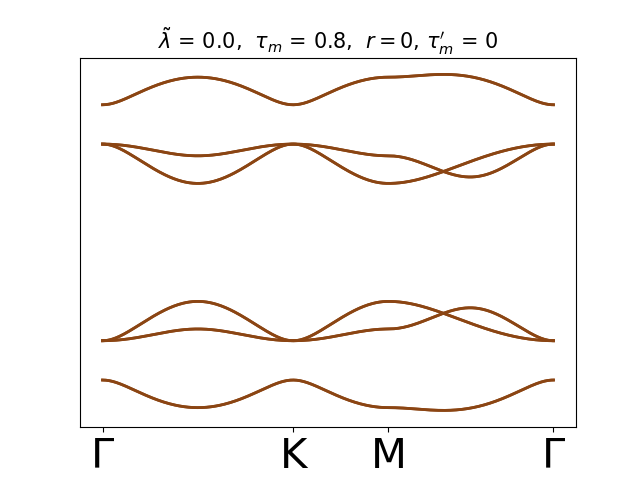} \qquad
    \includegraphics[scale=0.4]{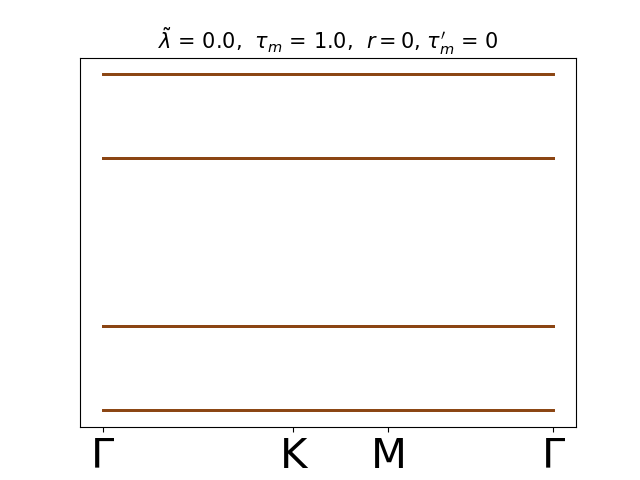}
	
    	\caption{Evolution of band structure along the $\tilde \lambda=0$ line of the phase diagram in Fig.~6 of the main text.}
	\label{fig_tsigma_to_tm_interpolation}
\end{figure}

Fig~\ref{fig_tsigma_to_tm_interpolation} shows the band structures at different points on the $\lambda=0$ line of the $\tau_m$-$\tilde \lambda$ phase diagram in Fig.~6a of the main text. At $\tau_m=0$ on this line, there are two flat bands, each of which are six-fold degenerate. As $\tau_m$ is increased, six 2-fold degenerate bands appear. Finally at $\tau_m=1$, there are four bands with the lowest and the top bands being 2-fold and the rest being 4-fold degenerate.

\subsubsection{The $\tau_m=0$ line}
\begin{figure}[h!]
	\includegraphics[scale=0.4]{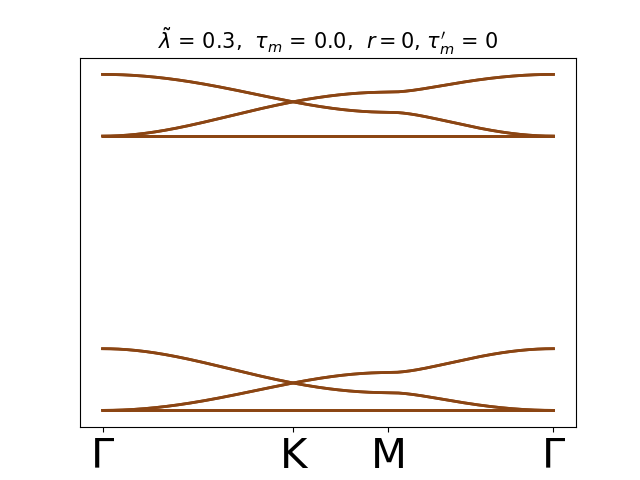} \qquad
	\includegraphics[scale=0.4]{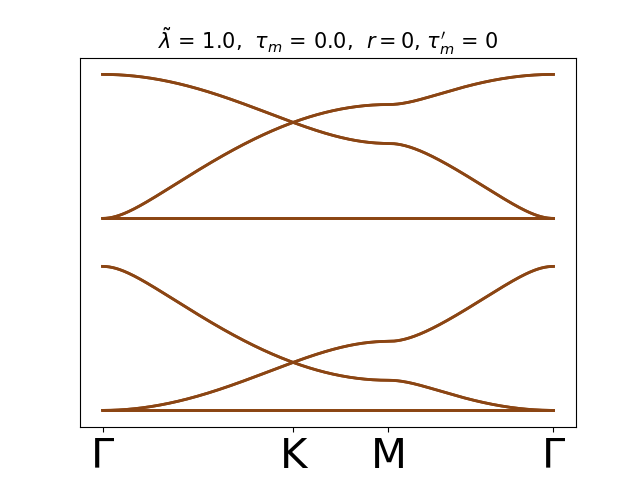}
	\\ \quad \\
	\includegraphics[scale=0.4]{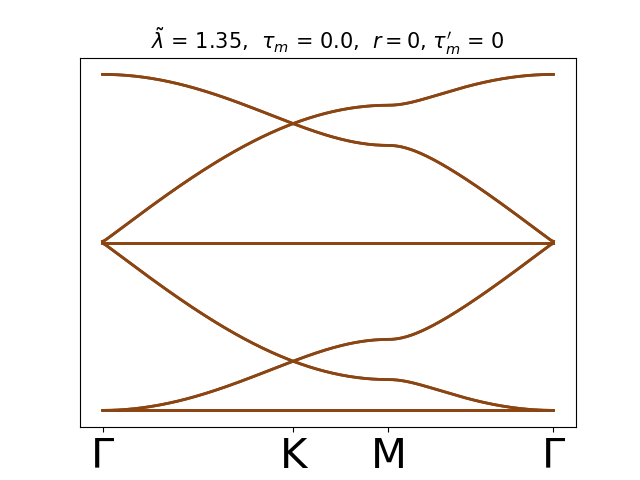} \qquad
	\includegraphics[scale=0.4]{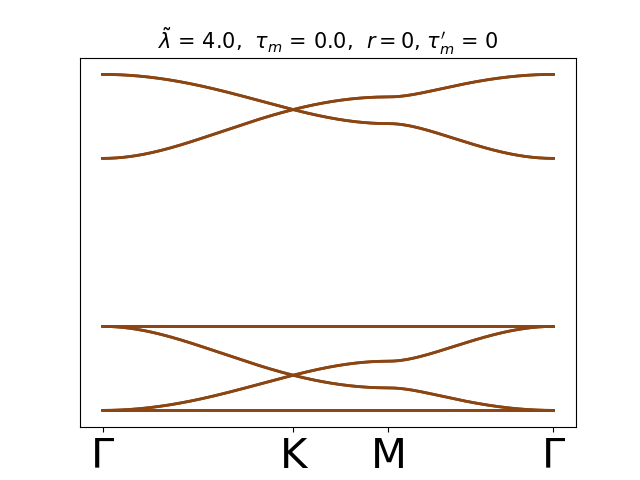}
	\\ \quad \\
	\caption{Evolution of band structure along the $\tau_m=0$ line of the phase diagram in Fig.~6 of the main text.}
	\label{fig_tsigma_for_different_soc}
\end{figure}

The energy spectrum along the $\tau_m=0$ line is shown in Fig.~\ref{fig_tsigma_for_different_soc}. The band structure for $\tilde \lambda =0$ is (cf. the $\tilde \lambda=0, \tau_m=0$ line in Fig.~\ref{fig_tsigma_to_tm_interpolation}) has two six-fold degenerate bands. As $\tilde \lambda$ is increased, six 2-fold degenerate bands appear, two of them being completely flat.

\subsubsection{The $\tau_m=1$ line}
\begin{figure}[h!]
	\includegraphics[scale=0.4]{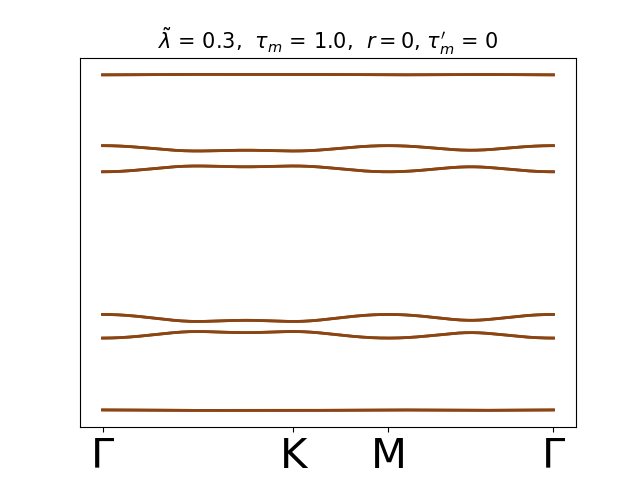} \qquad
	\includegraphics[scale=0.4]{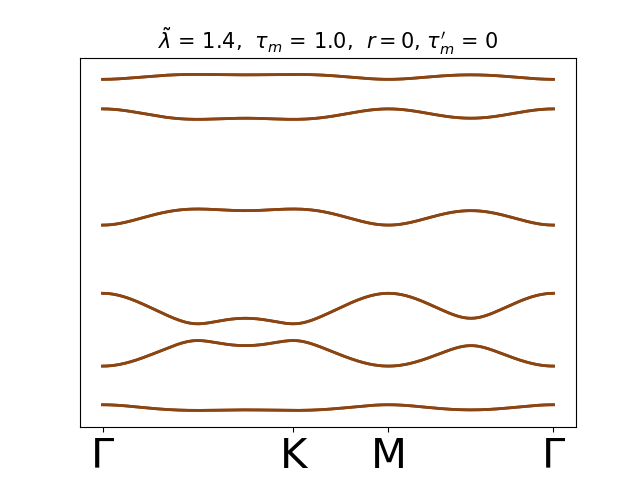} \\ \quad \\
	\includegraphics[scale=0.4]{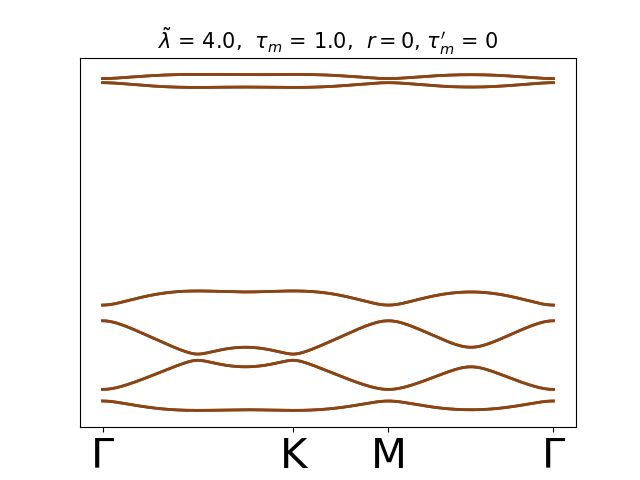} \qquad
	\includegraphics[scale=0.4]{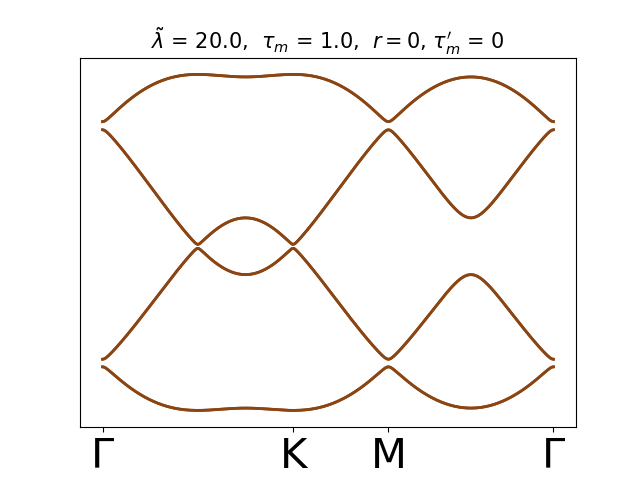} \qquad
	\caption{Band structures along the $\tau_m=1$ line of the phase diagram in Fig.~6 of the main text. All the bands are 2-fold degenerate. For $\tilde\lambda =20$, only four bands are shown leaving out higher energy $J=1/2$ orbitals.}
	\label{fig_tm_for_different_soc}
\end{figure}

The evolution of energy spectrum along the $\tau_m=1$ line is shown in Fig.~\ref{fig_tm_for_different_soc}. The band structure for $\tilde \lambda =0$ on this line is has four bands (cf. $\tilde 
\lambda=0, \tau_m=1$ in Fig.~\ref{fig_tsigma_to_tm_interpolation}). For large values of $\tilde \lambda$, the $J=3/2$ and the $J=1/2$ bands are separated. For $\tilde \lambda=20, \tau_m=1$, only the lowest $J=3/2$ bands are shown.

%%%%%%%%%%%%%%%%%%%%%%%%%%%%%%%

\section{Properties of gapped phases} \label{appen_gapped_phases}

\subsubsection{SU(8) Dirac theory at the $P_3$ point}

At $P_3$ point of the phase diagram in Fig.~6 and Fig.~11 of the main text, the low-energy effective theory is described by massless Dirac fermions with internal SU(8) symmetry. Below we sketch the derivation of the Dirac theory. The details can be found in  Ref.~\cite{mondal2023emergent}.

At the $P_3$ point, the form of the Hamiltonian in Eq.~5 of the main text, when projected to the low-energy $J=3/2$ orbitals, is given by
\begin{eqnarray}\label{eq_ham_at_P3}
    &&\mcl{H}_{P_3} = -\frac{\mcl{E}}{\sqrt{3}}\left(\sum_{\braket{ij}\in X-bonds} \psi^\dagger_i U_{X} \psi_j + \sum_{\braket{ij}\in Y-bonds} \psi^\dagger_i U_{Y} \psi_j  + \sum_{\braket{ij}\in Z-bonds} \psi^\dagger_i U_{Z} \psi_j \right).
\end{eqnarray}

Here, $\psi_i$ is a 4-component annihilation operator corresponding to the four $J=3/2$ orbitals at the site $i$. The $U_X, U_Y, U_Z$ are $4\times4$ Hermitian matrices, which are given by
\begin{eqnarray}\label{eq_u matrices}
    &&U_X = -\rho\Sigma_1 \\
    &&U_Y = -\rho\Sigma_2 \\
    &&U_Z = -\rho\Sigma_3.
\end{eqnarray}
Here, $\rho=+1$ for the phase diagram in Fig.~6a and $\rho=-1$ for Fig.~11. The $\Sigma_i$ (for $i=1,\cdots 15$) are the 15 linearly independent $4\times 4$ traceless Hermitian matrices which are given in Appendix.~C of Ref.~\cite{mondal2023emergent}, which are essentially the generators of a SU(4) group. Note, the Hamiltonian in Eq.~\ref{eq_ham_at_P3} has an internal SU(4) symmetry\cite{yamada_2018_emergent_SU(4)}.

Projecting this Hamiltonian in Eq~\ref{eq_ham_at_P3} to the two lowest two bands which touch linearly at the Fermi energy, we get the SU(8) symmetric Dirac Hamiltonian given by
\begin{eqnarray}\label{eq_su8 dirac Hamiltonian}
    \mcl{H}_{Dirac} = \rho v_F \int d^2\mbf{r}~\chi^\dagger(\mbf{r}) \left( -i\alpha_1 \partial_1 - i\alpha_2 \partial_2 \right) \chi(\mbf{r}),
\end{eqnarray}
with
\begin{eqnarray}
    &&\alpha_1 = \Sigma_0\otimes \tau_3\otimes\sigma_1 \\
    &&\alpha_2 = \Sigma_0\otimes \tau_0 \otimes \sigma_2.
\end{eqnarray}
Here, $v_F$ is the Fermi velocity, which is related to the gradient of the linearly dispersing bands at the Dirac points. $\chi(\mbf{r})$ is the 16-component Dirac spinor at the position $\mbf{r}$. Both $\tau_i$ and $\sigma_i$ (for $i=1,2,3$) are the Pauli matrices with $\tau_0$ and $\sigma_0$ being the $2\times 2$ identity matrix. The generators of the SU(8) symmetry are the set of 63 matrices given by 
\begin{eqnarray}
 \{\Sigma_0,\Sigma_i\} \otimes \{ \tau_3\sigma_0, \tau_1\sigma_2, \tau_2\sigma_2 \}, \quad\Sigma_i\otimes \tau_0\sigma_0
\end{eqnarray}
for $i=1,\cdots 15$.
%$\Sigma_i \otimes \{ \tau_3\sigma_0, \tau_1\sigma_2, \tau_2\sigma_2 \}$ (for $i=1,\cdots 15$). 

\subsubsection{The topological gapped phase}
On moving left from the $P_3$ point along the $\tilde \lambda =\infty$ line by reducing the value of $\tau_m$ from 1, the effective hopping Hamiltonian in the $J=3/2$ sector is given by 

\begin{eqnarray}\label{eq_gapped phase1 Haniltonian}
    &&\mcl{H}_{top}= -\frac{\mcl{E}}{\sqrt{3}}~\sum_{\braket{ij}} \psi^\dagger_i \tilde H_{ij} \psi_j, 
\end{eqnarray}
with 
\begin{eqnarray}
&&\tilde H_X = U_X + \left(1-\tau_m\right)\left[\frac{1}{3}\Sigma_0 + \frac{1}{6}\left( -\sqrt{3}\Sigma_4 + \Sigma_5 \right)\right] \nonumber\\
&& \tilde H_Y = U_Y  + \left(1-\tau_m\right)\left[\frac{1}{3}\Sigma_0 + \frac{1}{6}\left( \sqrt{3}\Sigma_4 + \Sigma_5 \right)\right] \nonumber\\
&&\tilde H_Z = U_Z  + \left(1-\tau_m\right)\left[\frac{1}{3}\Sigma_0 -\frac{1}{3}\Sigma_5)\right].
\end{eqnarray}

On projecting this Hamiltonian to the lowest two bands, we get the following effective low-energy Hamiltonian
\begin{eqnarray}
    \mcl{H}_{top} = \mcl{H}_{Dirac} + (1-\tau_m)\mcl{H}_{m}^{(1)} + (1-\tau_m) \mcl{H}^{\prime}.
\end{eqnarray}
Here, $\mcl{H}_{Dirac}$ is given by Eq.~\ref{eq_su8 dirac Hamiltonian}. The $\mcl{H}_m^{(1)}$ and $\mcl{H}'$ are given by
\begin{eqnarray}
    \mcl{H}_{m}^{(1)} = \int d^2\mbf{r}~  \chi^\dagger\left(\Sigma_1\tau_1\sigma_0 - \Sigma_2\tau_2\sigma_1 + \Sigma_3\tau_0\sigma_3  \right)\chi ,
\end{eqnarray}
and
\begin{align}
    \mathcal{H}'=\chi^\dagger \left(-i \partial_x \delta\alpha_x - i\partial_y \delta\alpha_y\right)\chi,
\end{align}
with 
\begin{align}
    \delta\alpha_x =& -\Sigma_3\tau_3\sigma_1 + \frac{1}{\sqrt{3}}\Sigma_{35}\tau_0\sigma_0 + \frac{\sqrt{3}}{\sqrt{2}}\Sigma_{35}\tau_0\sigma_3
\end{align}
and
\begin{align}
    \delta\alpha_y =& \frac{1}{\sqrt{6}}(\Sigma_{1}\tau_{1}\sigma_{2}-\Sigma_{2}\tau_{2}\sigma_{2}) + \frac{\sqrt{3}}{2\sqrt{2}}(\Sigma_{14}\tau_{1}\sigma_{1}-\Sigma_{24}\tau_{2}\sigma_{1}) - \frac{1}{2}(\Sigma_{14}\tau_{2}\sigma_{3}-\Sigma_{24}\tau_{1}\sigma_{3}) -\frac{1}{2\sqrt{2}}(\Sigma_{15}\tau_{1}\sigma_{1}+\Sigma_{25}\tau_{2}\sigma_{1}) \nonumber\\
     &+ \frac{1}{2\sqrt{3}}(\Sigma_{15}\tau_{2}\sigma_{3} + \Sigma_{25}\tau_{1}\sigma_{3})
\end{align}
The term $\mcl{H}_m^{(1)}$ is the ferro-quadrupolar quantum Hall mass listed in Ref.\cite{mondal2023emergent} whose edge modes are protected by the time reversal (TR) symmetry. Thus, the phase obtained by moving left from the $P_3$ point in the phase diagram in Fig.~6a is a $Z_2$ topological insulator. The presence of $\mcl{H}'$ does not change the topological character of this phase since this term does not break the TR symmetry and can be tuned to zero without closing the fermionic energy gap.

\subsubsection{The non-topological phase}
On moving down vertically from the $P_3$ point along the $\tau_m=1$ line, we encounter the non-topological gapped phase. This can be understood by doing a similar analysis as done for the previous gapped phase. For very large values of $\tilde\lambda$ and $\tau_m=1$, the effective Hamiltonian is given by 
\begin{eqnarray}
    \mcl{H}_{non-top} = -\mcl{E}\left(\sum_{\braket{ij}} \psi^\dagger_i U_{ij}\psi_j + \frac{1}{\lambda}\sum_{\braket{\braket{ij}}} \psi^\dagger_i \tilde{\tilde{H}}_{ij} \psi_j + h.c.\right) \nonumber\\
\end{eqnarray}

\begin{figure}
    \centering
    \includegraphics[scale=0.3]{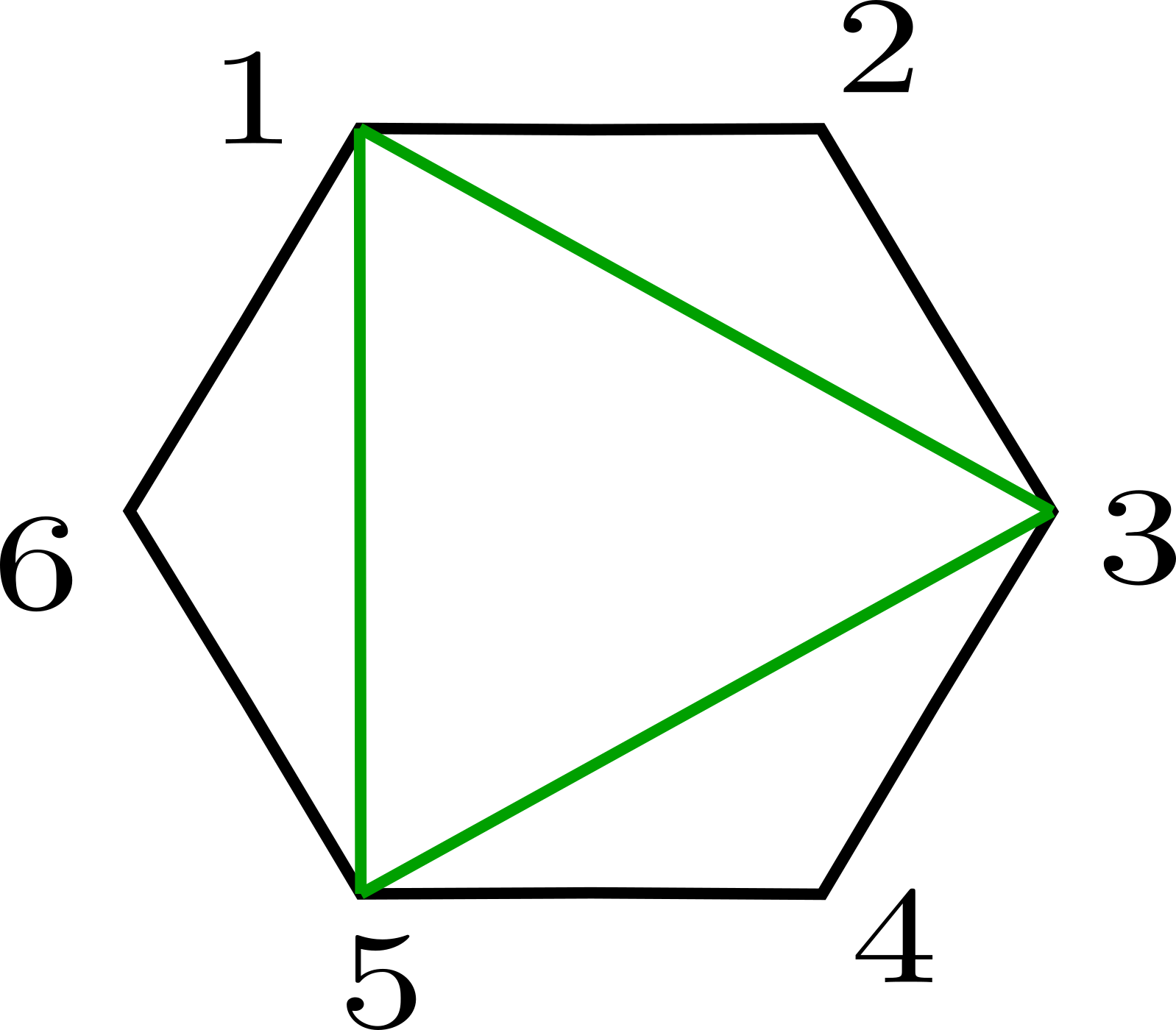}
    \caption{A single hexagon of a honeycomb lattice showing the three kinds of next nearest neighbour bonds (green bonds). The six sites are labelled with integers from 1 to 6.}
    \label{fig:nnn_hopping_gapped_phase}
\end{figure}

Where the $U_{ij}$ are the matrices defined in Eq.~\ref{eq_u matrices} and $\tilde{\tilde{H}}_{ij}$ are the hopping matrices on the next-nearest(NNN) bonds of the lattice. On the three kinds of NNN bonds shown in Fig~\ref{fig:nnn_hopping_gapped_phase}, the $\tilde{\tilde{H}}_{ij}$ matrices are given by 
\begin{align}
\tilde{\tilde{H}}_{13} = & -\frac{1}{2\sqrt{3}}\Sigma_1 + \frac{i}{6} \left(-\Sigma _{14}-\sqrt{3} \Sigma _{15}-\Sigma _{23}\right) \nonumber\\
\tilde{\tilde{H}}_{35} = & -\frac{1}{2\sqrt{3}}\Sigma_2 +\frac{i}{6}  \left(\Sigma _{13}-\Sigma _{24}+\sqrt{3} \Sigma _{25}\right)\nonumber\\
\tilde{\tilde{H}}_{51} = & -\frac{1}{2\sqrt{3}}\Sigma_3 +\frac{i}{6}  \left(-\Sigma _{12}+2 \Sigma _{34}\right)
\end{align}

On projecting this Hamiltonian to the lowest two bands near the Dirac points, we get the following low-energy theory
\begin{eqnarray}
    \mcl{H}_{non-top} = \mcl{H}_{Dirac} + \frac{1}{\lambda}\mcl{H}_{m}^{(2)} +  \frac{1}{\lambda}\mcl{H}^{\prime\prime}.
\end{eqnarray}
Here, $\mcl{H}_{Dirac}$ is the SU(8) symmetric Dirac Hamiltonian. The $\mcl{H}_m^{(2)}$ is given by
\begin{eqnarray}\label{eq_top mass 2}
    \mcl{H}_{m}^{(2)} = \int d^2x ~\chi^\dagger~\Sigma_{45}\tau_3\sigma_3 \chi
\end{eqnarray}
The $\mcl{H}_{m}^{(2)}$ is one of the topological masses proximate to the SU(8) Dirac semi-metal. The edge modes of this topological symmetry is protected by a U(1) symmetry which is generated by $\Sigma_{45}$. But the term $\mcl{H}^{\prime \prime}$ breaks this U(1) symmetry and destroys the edge modes. This explains the existence of the non-topological phase in the phase diagram. 

\subsubsection{Phase transition lines in $\tau_m$-$\tilde \lambda$ plane with $\rho$ = $\pm$ 1}

 On the phase transition line, the $P_1P_3$ line, between the two gapped phases shown in Fig.~6a of the main text, the lowest two bands touch each other linearly at the $\Gamma$ point as shown in inset~$VI$ of the same figure. One can find the low-energy theory for this point of the phase diagram by projecting the Hamiltonian to the two lowest band that touch at the $\Gamma$ point. The resultant theory is given by
 \begin{eqnarray}
     \mcl{H}_{\Gamma} = -iv_F \int d^2x ~\chi_{\Gamma}^\dagger\left( \tau_3\sigma_1 \partial_1 + \tau_0\sigma_2 \partial_2 \right)\chi_{\Gamma}.
 \end{eqnarray}

 Here, $\chi_{\Gamma}$ is a 4-component spinor which comes from the two-fold degenerate Dirac cone at the $\Gamma$ point. This Hamiltonian has an emergent SU(2) symmetry which is generated by $\{\tau_3\sigma_0, \tau_1\sigma_2, \tau_2\sigma_2\}/2$. 

 Similarly, on the $P_{3}P_{14}$ line of the phase of Fig.~11, the lower two bands touch each other linearly at the three M points. The low-energy theory at any point on this line is given by
 
\begin{eqnarray}
     \mcl{H}_{M} = i \int d^2x ~\chi_{M}^\dagger\left( v_x\mathbb{I}_{3\times 3} \otimes \tau_3\sigma_1 \partial_1 + v_y \mathbb{I}_{3\times 3}\otimes\tau_0\sigma_2 \partial_2 \right)\chi_{M}. 
 \end{eqnarray}

 Here, $\chi_M$ is a 12-component spinor that comes from the three two-fold degenerate Dirac cones at the three M points. $v_x$ and $v_y$ are the Fermi velocities along the two Cartesian directions. The values of these two numbers depend on the position on the phase transition line. $\mathbb{I}_{3\times 3}$ is the three dimensional identity matrix that acts on the space of the three M valleys. This Hamiltonian has an internal SU(6) symmetry which are generated by the set of Hermitian matrices given by
 \begin{eqnarray}
  \{\mathbb{I}_{3\times3},\Lambda_i\} \otimes \{\tau_3\sigma_0, \tau_1\sigma_2, \tau_2\sigma_2\}, \quad \Lambda_i\otimes \tau_0\sigma_0
 \end{eqnarray}
 %$\Lambda_i \otimes \{\tau_3\sigma_0, \tau_1\sigma_2, \tau_2\sigma_2\}$ (for $i=1,\cdots 8$),
 where the $\Lambda_i$ are the eight $3\times3$ Gell-Mann matrices that generate an SU(3).

\section{DFT Computed Magnetic Ground states}

As discussed in main text, one of the possible consequence of inclusion
of Coulomb correlation is to stabilize magnetism. To identify the magnetic ground states of the undimerized MX$_3$ compounds, DFT total energies for magnetic configurations, e.g 
non-magnetic(NM), ferromagnetic(FM), Neel-antiferromagnetic (AFM), zigzag-AFM (ZAFM), and stripe-AFM (SAFM) were calculated within the GGA+U+SOC formulation to take into account of the Coumlomb correlation in a mean-field way along with SOC, and compared. The Coulomb correlation is modeled
through supplemented Hubbard U correction and the Hund's coupling J to
account for the multi-orbital nature of the problem.

The calculated spin and orbital magnetic moments of the nine compounds are tabulated in Table I. 
As expected, the orbital moment shows an increasing trend in moving 
from Ti to Zr to Hf compounds, while the spin moment shows an decreasing trend. This is justified by increase of SOC in moving 3d to 4d to 5d transition metal series, and extended nature of the wavefunction in
moving from 3d to 4d/5d.

\begin{figure}
    \centering
    \includegraphics[width=0.7\columnwidth]{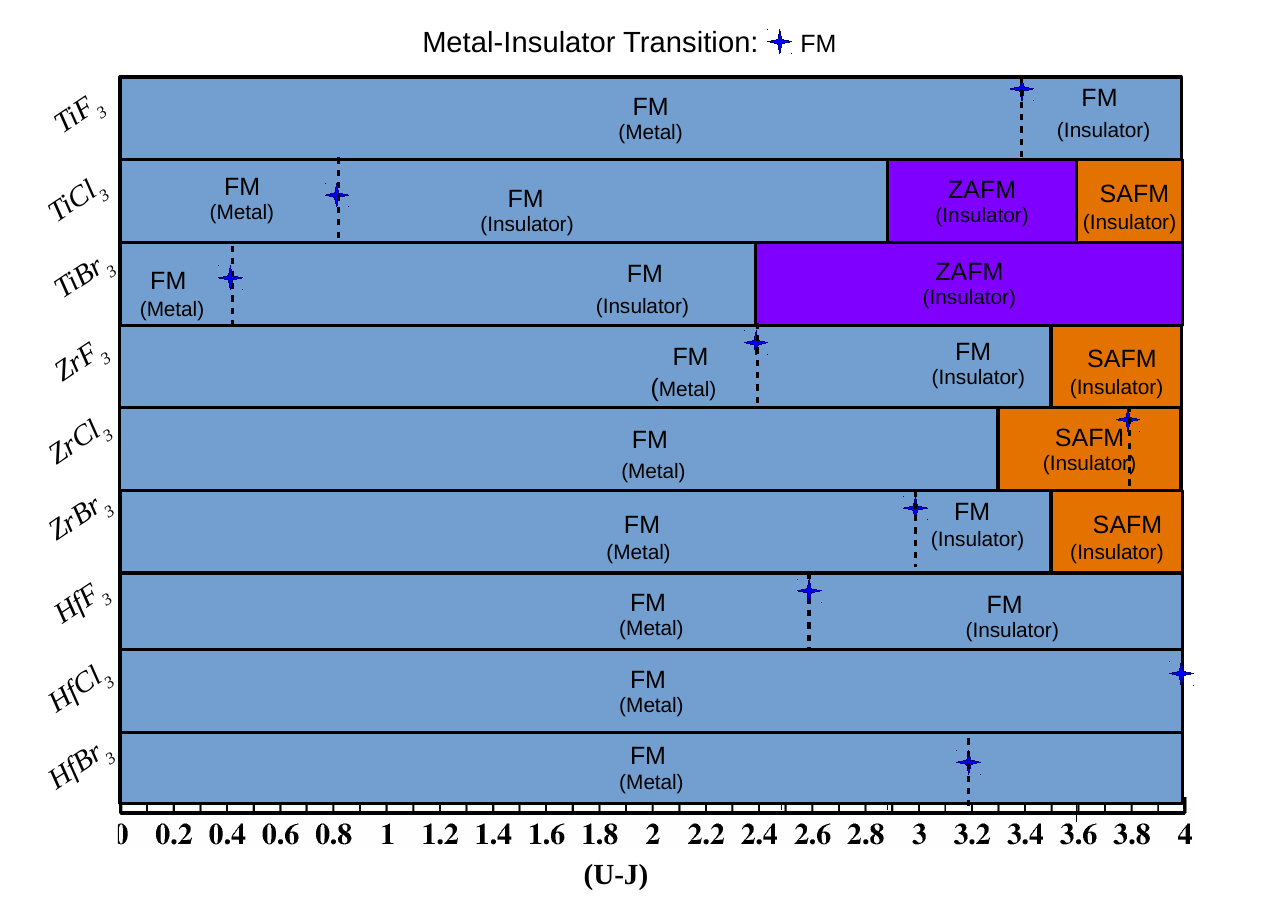}
    \caption{Magnetic Ground states for MX$_3$ for various different choices of (U-J) in eV.}
    \label{fig:magnetic_ground_state}
\end{figure}

The computed magnetic phase diagram, is shown in Fig~\ref{fig:magnetic_ground_state} in the plane of compounds versus
choice of (U-J) parameter. Marked are the lowest energy magnetic state
according to DFT total energy, their conducting properties estimated
from density of states plots. The metal-insulator transitions as
well as magnetic transitions are marked by boundaries.

First of all, we notice in a large part of the phase diagram, the
FM state is stabilized, with the exception of SAFM or ZAFM phases
at large value of (U-J) for Ti and Zr compounds. Although the (U-J)
is varied over a large range in the plot, for the realistic estimates
of U value for early transition metal like 3d Ti will be 3-4 eV, while
4d/5d Zr/Hf will be 1-2 eV. With estimated J value of 1 eV for 3d
transition metal and 0.4 eV for 4d/5d transition metal, this amounts
to (U-J) value of 2-3 eV for Ti compounds, and 0.6-1.6 eV for Zr/Hf
compounds. With this choice, undimerized TiF$_3$ turns to be FM
metal, while TiCl$_3$/TiBr$_3$ may exhibit stabilization of ZAFM phase. For 4d/5d Zr/Hf compounds, in undimerzed structure, the FM metallic phase wins over the AFM phases.

\begin{table} [h!]
\centering
\begin{tabular}{ | c c c | }
\hline
MX$_3$ & Orbital Moment( $\mu_\text{B}$) & Spin Moment( $\mu_\text{B}$) \\ [0.5ex]
\hline
TiF$_3$   &-0.061 &0.922   \\
TiCl$_3$  &-0.003 &0.938    \\
TiBr$_3$  &-0.008 &0.965     \\
ZrF$_3$   &-0.044 &0.722 \\
ZrCl$_3$  &-0.034  &0.661\\ 
ZrBr$_3$  &-0.026  &0.674 \\
HfF$_3$   &-0.132  &0.806 \\
HfCl$_3$  &-0.104  &0.626  \\
HfBr$_3$  &-0.111 &0.728\\
[1ex]
 \hline
\end{tabular}
 \caption{Orbital and Spin Magnetic Moments on Metal(M) site for MX$_3$, as calculated within GGA+U+SOC. (U-J) value was chosen to be 3 eV for
 Ti and 2 eV for Zr/Hf.}
 \label{table:orbital_moments}
\end{table}

\bibliography{ref}